\documentclass[twocolumn,english,aps,prb,twocolum,superscriptaddress,natbib,bibnotes,amsmath,amssymb,floatfix,groupedaddress,footinbib]{revtex4-2}
\usepackage[colorlinks=true,citecolor=blue,linkcolor=magenta]{hyperref}

\usepackage[addedmarkup=colored, authormarkupposition=left]{changes} 

\usepackage{soul}
\definechangesauthor[name={TJK}, color=red]{TJK}

\usepackage[utf8]{inputenc}
\usepackage[english]{babel}
\usepackage{amsmath,amsfonts,amssymb}
\usepackage[T1]{fontenc}
\usepackage{xurl}

\usepackage{amsmath}
\usepackage{siunitx}
\usepackage[version=4]{mhchem}
\usepackage{upgreek}

\usepackage{amsfonts}
\usepackage{amssymb}

\usepackage{epstopdf}
\usepackage{graphicx}
\graphicspath{{./Figures/}}

\begin{document}

\title{Full C- and L-band tunable erbium-doped integrated lasers via scalable manufacturing}

\author{Xinru Ji$^{1,2}$, 
				Xuan Yang$^{1,2}$, 
				Yang Liu$^{1,2}$,
			    Zheru Qiu$^{1,2}$, 
			    Grigory Lihachev$^{1,2}$, 
			    Simone Bianconi$^{1,2}$,
			    Jiale Sun$^{1,2}$, 
                Andrey Voloshin$^{1,2}$,
                Taegon Kim$^{3}$,
                Joseph C. Olson$^{3}$,
				and Tobias J. Kippenberg$^{1,2\dag}$}
\affiliation{
$^1$Institute of Physics, Swiss Federal Institute of Technology Lausanne (EPFL), CH-1015 Lausanne, Switzerland\\
$^2$Center for Quantum Science and Engineering, Swiss Federal Institute of Technology Lausanne (EPFL), CH-1015 Lausanne, Switzerland\\
$^3$Varian Semiconductor, Applied Materials, Gloucester, MA 01930, United States
}

\maketitle

\noindent\textbf{
Erbium (Er) ions are the gain medium of choice for fiber-based amplifiers and lasers, offering a long excited-state lifetime, slow gain relaxation, low amplification nonlinearity and noise, and temperature stability compared to semiconductor-based platforms.
Recent advances in ultra-low-loss silicon nitride (Si$_3$N$_4$) photonic integrated circuits, combined with ion implantation, have enabled the realization of high-power on-chip Er amplifiers and lasers with performance comparable to fiber-based counterparts, supporting compact photonic systems.
Yet, these results are limited by the high (2~MeV) implantation beam energy required for tightly confined Si$_3$N$_4$ waveguides (700~nm height).
The extended implantation time, large ion fluences, and high energy have thus far prevented volume manufacturing of Er-doped photonic integrated circuits.
Here, we overcome these limitations and demonstrate the first fully wafer-scale, foundry-compatible Er-doped photonic integrated circuit-based tunable lasers.
Using 200 nm-thick Si$_3$N$_4$ waveguides, we reduce the ion beam energy requirement to below 500~keV, enabling efficient wafer-scale implantation with an industrial 300 mm ion implanter.
The reduced implantation energy marks a crucial advance toward scalable production of Er-doped photonic devices.
Meanwhile, the increased optical mode area of low-confinement Si$_3$N$_4$ waveguides significantly enhances laser performance and output power.
We demonstrate a laser wavelength tuning range of 91~nm, covering nearly the entire optical C- and L-bands, with fiber-coupled output power reaching 36~mW and an intrinsic linewidth of 95~Hz.
The temperature-insensitive properties of erbium ions allowed stable laser operation up to \SI{125}{\degreeCelsius} and lasing with less than 15~MHz drift for over 6 hours at room temperature using a remote fiber pump.
The fully scalable, low-cost fabrication of Er-doped waveguide lasers opens the door for widespread adoption in coherent communications, LiDAR, microwave photonics, optical frequency synthesis, and free-space communications.}

\begin{figure*}[t!]
\centering
\includegraphics[width=\textwidth]{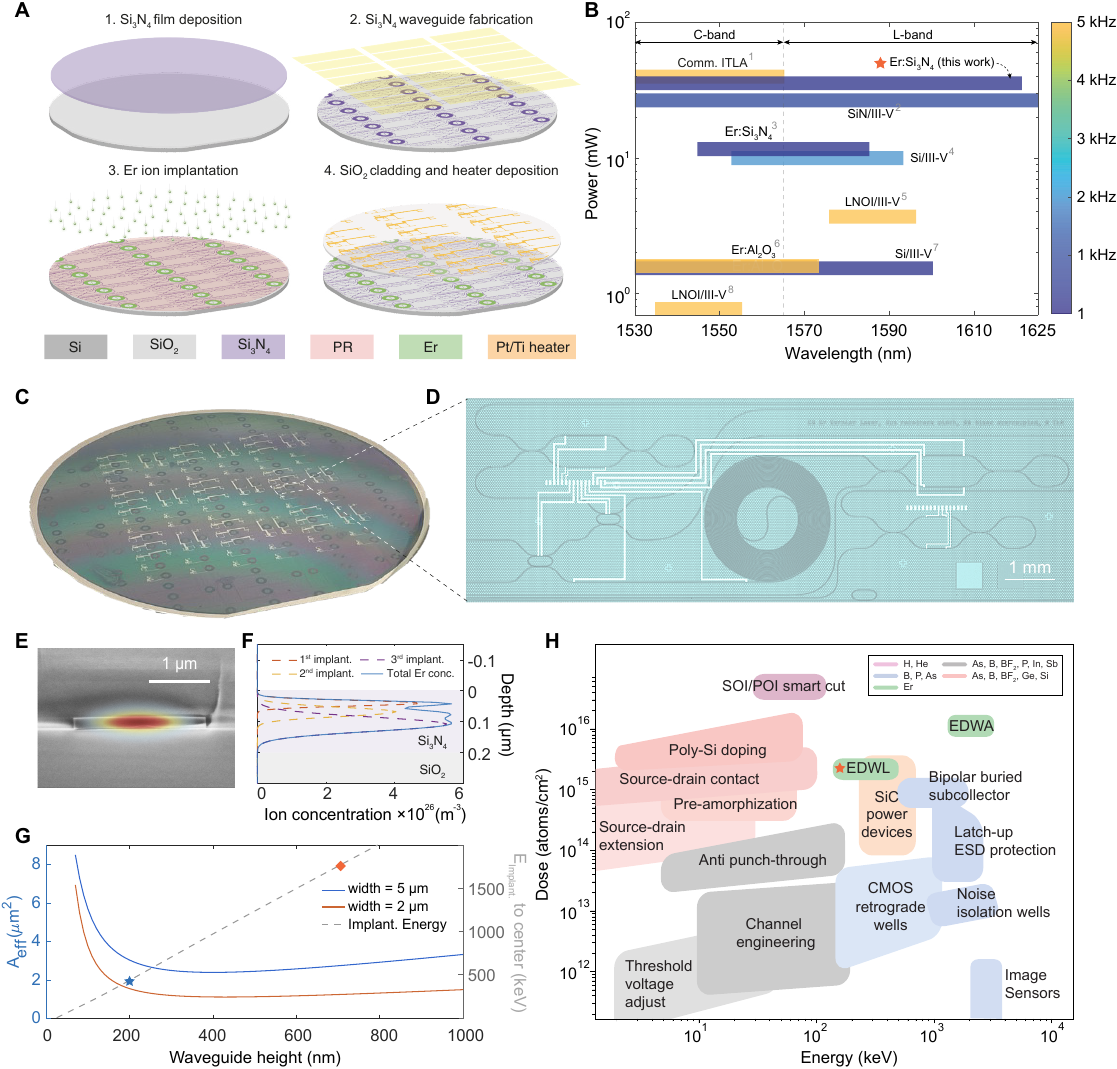}
\caption{
\footnotesize
\textbf{Wafer-scale manufacturing of Er-doped Si$_3$N$_4$ (Er:Si$_3$N$_4$) photonic integrated circuits.}
(\textbf{A}) Schematic of key processes for wafer-scale fabrication of Er:Si$_3$N$_4$ PICs, including 4-inch Si$_3$N$_4$ photonic wafer fabrication, Er ion implantation, and micro-heater deposition.
A 6~$\mathrm{\upmu}$m-thick photoresist mask (AZ 15XT), patterned via laser writing, enables wafer-scale selective Er ion implantation to Si$_3$N$_4$ spiral waveguides serving as gain sections.
(\textbf{B}) Comparison of key performance metrics for state-of-the-art integrated tunable lasers, including wavelength tuning range, output power, and intrinsic linewidth.
References for the compared devices:
Comm. ITLA (1): \href{https://www.lumentum.com/en/products/ultra-narrow-linewidth-nano-integrable-tunable-laser-assembly-nitla}{Lumentum ITLA};
SiN/III-V (2): \cite{guo2022hybrid};
Er:Si$_3$N$_4$ (3): \cite{liu_fully_2023};
Si/III-V (4): \cite{tran2019ring};
LNOI/III-V (5): \cite{li2022integrated};
Er:Al$_2$O$_3$ (6): \cite{li2018monolithically};
Si/III-V (7): \cite{morton2022integrated};
LNOI/III-V (8): \cite{op2021iii}.
(\textbf{C}) Optical image of a 4-inch wafer of Er:Si$_3$N$_4$ integrated tunable lasers following the fabrication processes outlined in (\textbf{A}).
(\textbf{D}) Optical image of a single Er:Si$_3$N$_4$ integrated tunable laser with a footprint of 0.4$\times$1.0 cm$^2$.
(\textbf{E}) Scanning electron microscope image of a 200-nm-thick Er:Si$_3$N$_4$ waveguide overlaid with a simulated fundamental transverse electric (TE) optical mode.
(\textbf{F}) Simulated erbium ion distribution along the vertical direction of a 200-nm-thick Si$_3$N$_4$ waveguide, based on three consecutive ion implantation steps with a maximum beam energy of 480~keV, optimized for overlap between Er ions and the optical mode.
(\textbf{G}) Simulated effective optical mode areas of the fundamental TE mode and required implantation beam energy to reach the waveguide center, as a function of the Si$_3$N$_4$ waveguide thickness.
The effective optical mode areas of 2.1~${\upmu}$m-wide \cite{Liu2024}, 700~nm-thick waveguides and 5~$\mathrm{\upmu}$m-wide, 200~nm-thick waveguides (this work) are shown for comparison.
(\textbf{H}) Implantation energy and dose of commonly used ion implantation processes in the microelectronics industry \cite{rubin2003ion, felch2013ion}, compared to the energy and dose required for previous implementations of Er-doped waveguide amplifier (EDWA) \cite{liu_photonic_2022} and Er-doped waveguide laser (EDWL) \cite{Liu2024}, as well as for the EDWL implementation presented in this work (marked with red star).
}
\label{Fig:1}
\end{figure*}

Erbium-doped fiber lasers (EDFLs) \cite{barnes_high-quantum-efficiency_1989,suzuki_8_1989,chow_multiwavelength_1996} are regarded as the benchmark for lowest laser phase noise, and play a key role in a wide range of applications such as distributed fiber sensing \cite{soriano-amat_time-expanded_2021,zadok_random-access_2012}, gyroscopes \cite{dix-matthews_point--point_2021}, free-space communications \cite{dix-matthews_point--point_2021}, and optical frequency metrology \cite{xu_recent_2013}.
Their advantages are attributed to the unique properties of Er ions, including a long excited-state lifetime, slow gain dynamics, temperature stability, and low noise figure \cite{desurvire1995erbium}.
However, their bulky size, high cost from manual assembling and complexity have relegated Er-doped fibers predominantly to laboratory environment and limited industrial use where footprint is less critical.
The realization of integrated lasers based on Er-doped photonic waveguides --- using Er ions as the same gain basis as in fiber lasers --- offers the potential for device miniaturization, fiber laser coherence, and temperature insensitivity in a monolithic architecture.
Nevertheless, hybrid Er-based integrated lasers have historically underperformed compared to commercial fiber lasers in terms of intrinsic linewidth and output power \cite{
kringlebotn1994er,snitzer_rare-earth-doped_2001,bastard_glass_2003, bernhardi2010ultra,cranch2011fundamental, belt2013arrayed,purnawirman_ultra-narrow-linewidth_2017, fu2017review,xiao_single-frequency_2021}, until recently:
Direct Er ion implantation into ultra-low-loss Si$_3$N$_4$ waveguides has enabled the realization of Er-doped waveguide amplifiers (EDWAs) with optical gain exceeding 30 dB and output power over 100~mW \cite{liu_photonic_2022}.
This advancement has further led to the development of Er-doped waveguide lasers (EDWLs) \cite{Liu2024}, which match the coherence of commercial Er-doped fiber lasers while surpassing them in tunability.
These achievements were realized in 700 nm-thick, tightly confined Si$_3$N$_4$ waveguides.
However, such thick Si$_3$N$_4$ waveguides require high implantation energies (up to 2~MeV) to achieve sufficient ion penetration for optimal mode overlap with the optical field.
This significantly limits system scalability, as it demands specialized implantation equipment with small area coverage (typically <2~$\times$ 2~cm$^2$), low beam current, long processing times, and introduces challenges such as heating and waveguide deformation from high-energy ion beams~\cite{liu_photonic_2022}.

Here, we overcome these challenges and demonstrate the first C+L band erbium-based tunable waveguide lasers fabricated via fully wafer-scale manufacturing.
This is realized through a low-confinement Si$_3$N$_4$ platform with a 200 nm waveguide thickness, which substantially reduces the required ion beam energy to below 500~keV.
This approach ensures compatibility with standard industrial implanters for 8- to 12-inch wafers (Fig. \ref{Fig:1}H), improving cost-efficiency, throughput, and fabrication scalability of Er-doped Si$_3$N$_4$ (Er:Si$_3$N$_4$) devices using established semiconductor protocols~\cite{rubin2003ion,felch2013ion}.
It reduces Er implantation time from tens of hours for a 2$\times$2 cm$^2$ area to tens of minutes for 12-inch wafers (Supplementary Note 1), minimizes waveguide deformations (Fig. \ref{Fig:1}E) compared to high-energy methods~\cite{liu_photonic_2022}, and bridges research-grade and industrial-scale fabrication.
A detailed comparison of Er ion implantations in high- and low-confinement Si$_3$N$_4$ waveguides is provided in Supplementary Note 1.
Furthermore, the low-confinement design mitigates optical nonlinear effects that degrade laser performance.
Collectively, we present fully hybrid integrated erbium-based lasers with a compact 0.4$\times$1.0 cm$^2$ footprint, a 91~nm wavelength tuning range spanning nearly the entire C- and L-bands, and fiber-coupled output power up to 36~mW and an intrinsic linewidth of 95~Hz.
This outperforms the power and tuning range of commercial Er-doped fiber lasers and state-of-the-art III-V based integrated lasers (Fig. \ref{Fig:1}B, \cite{guo2022hybrid,tran2019ring,li2022integrated,li2018monolithically,
morton2022integrated,op2021iii}), paving the way for scalable, cost-effective manufacturing of rare-earth-doped photonic integrated circuits, providing affordable high-coherence light sources for a wide range of applications.

\pretolerance=5000
\begin{figure*}[t!]
\centering
\includegraphics[width=\textwidth]{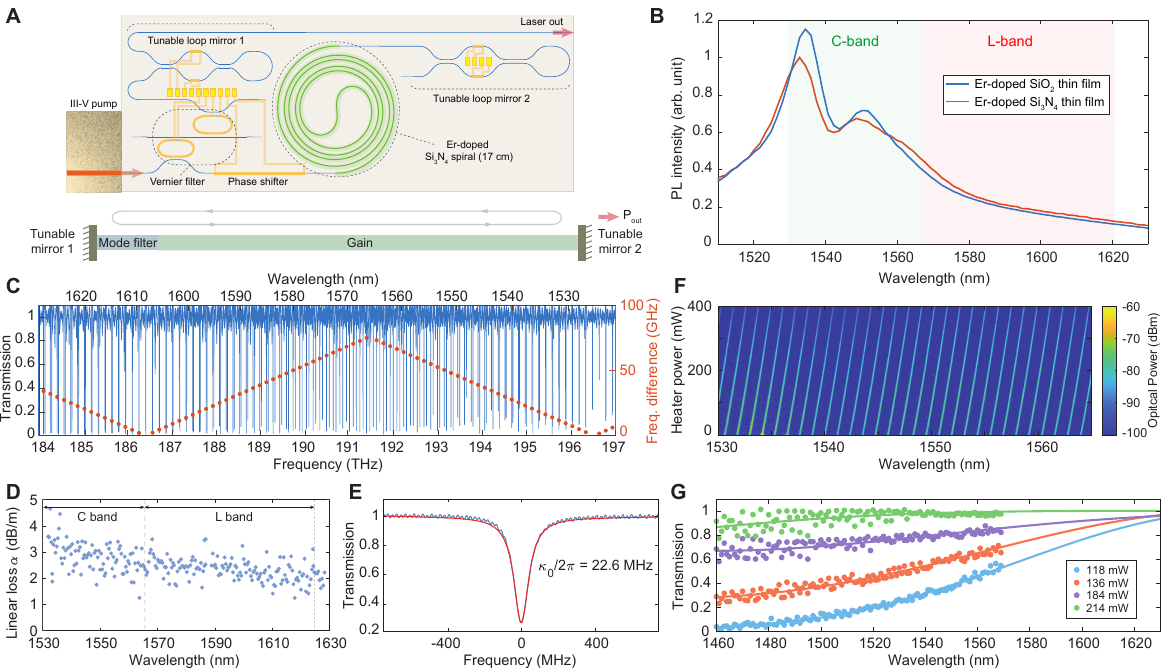}
\caption{
\footnotesize
\textbf{Characterization of volume-manufactured Er:Si$_3$N$_4$ integrated laser components.}
(\textbf{A}) Schematic of a hybrid integrated tunable Er laser featuring a microresonator-based Vernier filter for wavelength tuning, broadband loop mirrors for tunable mode reflection, and an Er:Si$_3$N$_4$ gain section.
(\textbf{B}) Measured photoluminescence (PL) spectra of Er-doped Si$_3$N$_4$ and \ce{SiO2} films, pumped by a 520 nm argon laser to excite the $^{4}I_\mathrm{15/2}$ state via decay from higher energy levels, free from in-band stimulated emission.
Both PL intensities are area-normalized and scaled for clarity.
(\textbf{C}) Broadband optical transmission measured at the drop-port of two Vernier microresonators with slightly differing FSRs.
The resulting Vernier spacing is approximately 10~THz, corresponding to an 85~nm wavelength range.
(\textbf{D}) Propagation loss of passive Si$_3$N$_4$ waveguides, obtained from the intrinsic linewidths of a 50 GHz FSR microring resonator with 5~$\upmu$m width and 200~nm height.
(\textbf{E}) A measured optical resonance (blue) with a fitted intrinsic linewidth $\kappa_0/2 \pi$ of 22.6~MHz (red).
(\textbf{F}) Thermo-optical tuning characterization of the integrated heater on a Vernier microresonator.
The optical spectrum illustrates the wavelength tuning of filtered amplified spontaneous emission (ASE) as the heater power increases from 0 to 400 mW.
(\textbf{G}) Measured (dots) and simulated (solid lines) transmission and of a broadband tunable loop mirror.
The transmission response is thermally tuned by a microheater.
}
\label{Fig:2}
\end{figure*}

\subsection*{Wafer-scale ion implantation and manufacturing of Er-doped photonic integrated circuits}

The wafer-scale manufacturing process of the EDWL starts with fabricating ultra-low loss Si$_3$N$_4$ photonic integrated circuits (Fig.~\ref{Fig:1}A).
The Si$_3$N$_4$ waveguides are formed from 200~nm low-pressure chemical vapor deposition (LPCVD) Si$_3$N$_4$ films on Si wafers with a 8~$\upmu$m wet-oxidized SiO$_2$ layer.
Post-deposition annealing at 1200$^\circ$C eliminates hydrogen effusion \cite{jiang2016effect}, minimizing absorption in the 1500–1538~nm range.
Waveguides are patterned with deep ultraviolet (DUV) stepper lithography and dry-etched to achieve smooth and vertical sidewalls.
Details on ultra-low loss Si$_3$N$_4$ PICs fabrication are provided in the Methods and Supplementary Note 2.

Prior to the Er ion implantation, a 6~$\upmu$m-thick photoresist layer (AZ 15nXT) is applied to selectively expose regions for Er implantation while shielding passive areas.
Er ion implantation is performed on passive Si$_3$N$_4$ waveguides using a commercial VIISta HE ion implanter, capable of processing areas up to 300~mm in diameter and supporting simultaneous implantation into four 4-inch wafers used in this work.
To maximize overlap between the erbium ions and the fundamental TE optical mode while minimizing parasitic upconversion, three consecutive implantation steps are carried out at energies of 480, 270, and 130~keV, achieving a homogeneous ion distribution.
The optimized ion fluences for each step are 3.2$\times$10$^{15}$, 1.5$\times$10$^{15}$, and 1.1$\times$10$^{15}$~cm$^{-2}$, respectively, resulting in a total implantation time of 100 minutes (details in Supplementary Note 1).
Figure \ref{Fig:1}F shows the simulated erbium ion concentration profile using the Monte Carlo program package 'Stopping and Range of Ions in Matter' (SRIM \cite{srim}), with a maximum transporting depth over 100~nm, achieving an overlap of $\Gamma \approx 30\%$.

After ion implantation, the wafer is annealed at 1000$^\circ$C for 1 hour in N$_2$ to optically activate erbium ions and heal implantation-induced defects.
A 3~$\upmu$m hydrogen-free, low-loss \ce{SiO2} cladding is deposited using SiCl$_4$ and O$_2$ precursors \cite{qiu_hydrogen_free_2023}. Platinum (Pt) and titanium (Ti) microheaters are then fabricated for thermo-optic tuning of the Vernier filter and loop mirrors.

\subsection*{Hybrid integrated erbium-based Vernier laser}

The erbium-based laser (Fig.~\ref{Fig:2}A) comprises a linear optical cavity with an Er-doped gain section between two tunable loop mirrors, and a Vernier-mode filter formed by two microresonators with slightly different free spectral ranges (FSRs).
The cavity maintains a uniform height of 200~nm and an optical longitudinal mode spacing of 300~MHz.
Figure \ref{Fig:2}B presents the photoluminescence (PL) spectra of Er-implanted Si$_3$N$_4$ and SiO$_2$ thin films, excited by a 520~nm argon laser which populates the $^{4}I_\mathrm{15/2}$ state via non-radiative decay from higher energy levels, avoiding in-band stimulated emission.
The area-normalized and scaled PL intensities indicate that the profile and bandwidth of Er ions in Si$_3$N$_4$ closely resemble those in SiO$_2$, a representative optical fiber medium.
The relative intensities of the two main PL peaks are attributed to variations in the local crystal field.

The erbium ions in the integrated laser platform are optically pumped via two approaches: edge coupling with a 1480~nm III-V laser diode (3SP 1943LCV1) or remote pumping using a high-power fiber-coupled module (QPhotonics QFBGLD-1480-500).
The edge couplers have a height of 200~nm and widths tapering from 0.42~$\upmu$m at the input and 0.5~$\upmu$m at the output.
Simulations indicate an insertion loss of 0.97~dB for the input coupler when edge-coupled to the 1480~nm pump diode and 0.56~dB when butt-coupled with UHNA-7 fibers.

An intra-cavity Vernier ring filter comprising two cascaded add-drop racetrack microresonators (Fig. \ref{Fig:2}A) is implemented to achieve precise laser mode selection.
To ensure single-mode operation and minimize transmission losses to higher-order transverse modes, both resonators feature a waveguide width of 2~$\upmu$m and a 3~dB bandwidth narrower than the laser cavity longitudinal mode spacing.
The microresonators designed with FSRs of 144~GHz and 142~GHz result in a Vernier FSR of 10.224~THz, matching the wavelength span of the optical C+L bands.
The broadband transmission measured at the drop-port of two Vernier resonators (Fig. \ref{Fig:2}C), along with the frequency differences between neighboring resonances (in red), confirms a Vernier filter spacing of $\sim$10~THz, covering the major Er emission band.
Figure~\ref{Fig:2}D presents the propagation loss $\alpha$ of a passive microring resonator with a 50~GHz FSR and 5~$\upmu$m width, measured using scanning laser spectroscopy at transverse electric (TE) polarization.
Figure~\ref{Fig:2}E shows a representative optical resonance with a fitted intrinsic linewidth of $\kappa_0 / 2\pi = 22.6$~MHz.
$\alpha$ (dB/m) is derived from the fitted intrinsic resonance loss $\kappa_0 / 2\pi$ using $\alpha = 10 \log_{10}(\mathrm{e}) \cdot n_\mathrm{g} \kappa_0 / c$, where the group index $n_\mathrm{g} = 1.8$.
The waveguide propagation loss $\alpha$ varies between 1.5 and 4~dB/m across the C- and L-bands, with wavelength dependence primarily attributed to Rayleigh scattering at the waveguide sidewalls.

Pt/Ti microheaters are employed to tune the laser emission wavelength by aligning the peak transmission of the Vernier filter with a cavity longitudinal mode.
The tuning efficiency of the microheater is characterized in Fig. \ref{Fig:2}F, where the optical spectrum map demonstrates wavelength tuning of filtered amplified spontaneous emission (ASE) measured at the drop-port of the Vernier resonator.
Single-mode linear tuning is observed as the heater power increases from 0 to 400~mW, achieving a tuning efficiency of 620.6 GHz/W. 

The tunable broadband cavity reflectors use a looped Mach-Zehnder interferometer (MZI) structure.
Directional couplers within the MZI regulate power distribution between the two arms, determining the mirror output.
Phase tuning is achieved via microheaters, which introduce precise phase shifts to flexibly control the broadband transmission and reflection characteristics of the loop mirrors (Fig.~\ref{Fig:2}G).
A comprehensive analysis of the loop mirror design and tuning performance of fabricated devices is provided in Supplementary Note 3, demonstrating full transmission and reflection under specific bias conditions.

\subsection*{Laser wavelength tuning and emission coherence}
\begin{figure*}[t!]
\centering
\includegraphics[width=\textwidth]{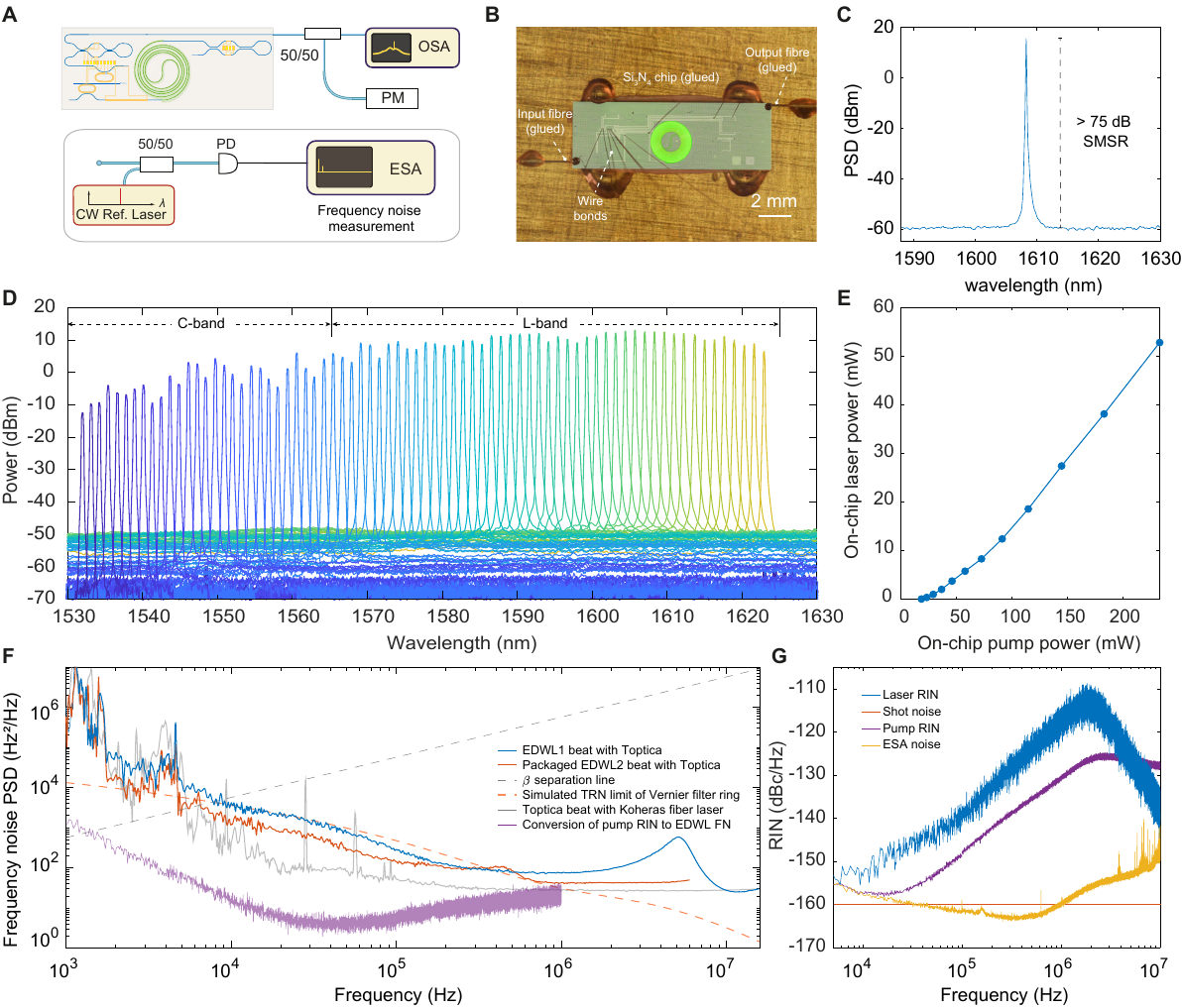}
\caption{
\footnotesize
\textbf{Performance of EDWLs in optical C + L bands.} 
(\textbf{A}) Experimental setup for laser wavelength tuning demonstration.
OSA: optical spectrum analyzer, PM: power meter, PD: photodiode, ESA: electric spectrum analyzer, CW Ref. laser - Toptica CTL.  
(\textbf{B}) Optical image of a fully packaged integrated Er-based tunable laser assembly.
(\textbf{C}) Optical spectra of lasing with 36 mW output power measured in fiber, under remote pumping.
The side mode suppression ratio (SMSR) is larger than 75 dB (Device ID: \texttt{D125\_43\_F5\_C2}).
(\textbf{D}) Optical spectrum of single-mode lasing across the optical C + L bands.
Wavelength tuning is achieved by adjusting Vernier filter and loop mirror reflection with microheaters.
Er ions are bidirectionally pumped using a fiber array with spacing matched to the separation between the input coupler and the mode-selective loop mirror, which transmits 1480~nm pump light and reflects the 1550~nm lasing (Device ID: \texttt{D125\_04\_F9\_C3}).
The resolution bandwidth of the optical spectrum analyser (OSA) is set to 0.2 nm.
(\textbf{E}) Dependence of lasing power on pump power, both measured on-chip, showing a 17.5~mW threshold and $\sim$24$\%$ slope efficiency (Device ID: \texttt{D125\_04\_F4\_C3}).
(\textbf{F}) Laser frequency noise of free-running (blue) and packaged (red) EDWLs, using heterodyne detection with a reference ECDL laser.
The frequency noise of the reference ECDL, measured via heterodyne detection with a stable fiber laser, is shown in grey.
The $\beta-$separation line is shown in dashed grey.
Simulated thermal refractive noise (TRN) limit for single Vernier filter ring is shown in dashed red.
Estimated transduction from pump laser RIN to EDWL frequency noise is presented in violet.
EDWL 1 device ID: \texttt{D125\_04\_F4\_C3}.
EDWL 2 device ID: \texttt{D125\_43\_F2\_C2}.
(\textbf{G}) Measured RIN of EDWL1 (blue), 1480 nm pump laser (violet), ESA noise floor (yellow), and shot noise limit (red).
}
\label{Fig:3}
\end{figure*}

We investigated the EDWL lasing performance and coherence using the experimental setups illustrated in Fig. \ref{Fig:3}A. 
To minimize external disturbances and validate laser performance, we performed photonic packaging in a custom 14-pin butterfly package (Fig. \ref{Fig:3}B).
A Peltier element, a thermistor and all microheaters are connected to butterfly pins using wire bonding.
Erbium ions in the packaged device are optically excited from $^{4}I_\mathrm{15/2}$ to $^{4}I_\mathrm{13/2}$ level by intra-band pumping from a 1480~nm multi-mode laser diode (QPhotonics QFBGLD-1480-500, $>$4~nm spectral width near 1480~nm) with 400~mW nominal power at 1.5~A driving current.
The laser input and output waveguides were edge-coupled and glued using cleaved UHNA-7 optical fibers spliced to SMF-28 fiber pigtails, resulting in insertion losses of 1.62~dB at the input and 1.28~dB at the output at 1550~nm.
The optical spectrum in Fig. \ref{Fig:3}C shows single-mode lasing at 1608~nm with 36~mW output power in fiber and a 75~dB side mode suppression ratio (SMSR) with 0.5~nm analyzer resolution bandwidth.

Figure \ref{Fig:3}D demonstrates the off-chip single-mode laser tuning across a broad wavelength range, from 1530 nm to 1621.3 nm.
In this measurement, the Er:Si$_3$N$_4$ waveguides were bidirectionally pumped via a fiber array designed to match the spacing between the input coupler and the mode-selective loop mirror, which transmits 1480 nm pump light while reflecting the 1550 nm lasing wavelength (Supplementary Note 3).
Such pumping configuration sufficiently excites Er ions to the $^{4}I_{13/2}$ state across the entire gain waveguide.
Mode-hop-free laser wavelength tuning is realized by adjusting the heater power on one microresonator in the Vernier filter, aligning distinct cavity modes with the filter passband.
To maximize output power at the target wavelength, the phase shifter inside the laser cavity and loop mirrors were adjusted during the optimization process.
The discrete lasing in Fig. \ref{Fig:3}D corresponds to the FSR of the Vernier microresonator.
Continuous tuning is achievable with simultaneous tuning of both microresonators.

This laser tuning exceeds the capabilities demonstrated in previous integrated laser systems \cite{liu_fully_2023,tran2019ring,li2022integrated,li2018monolithically,morton2022integrated,op2021iii},  achieving performance comparable to that of benchtop EDFLs optimized for EDF length and intra-cavity loss~\cite{dong2003linear}.
Nevertheless, a decrease in power is observed in the C-band compared to the L-band.
This is attributed to increased waveguide losses caused by O-H absorption in the SiO$_2$ cladding, resulting from cross-contamination of hydrogen in the SiO$_2$ deposition chamber.
O-H bonds exhibit strong overtone absorption near 200~THz, reducing the effective gain and lasing power of the EDWL in the C-band, as confirmed by simulations in Supplementary Note 4.
In contrast, the L-band output is less impacted due to the weaker absorption tail at longer wavelengths.
Theoretical simulations in Supplementary Note 4 indicate that, in the absence of O-H absorption and with waveguide loss following the fitted $\alpha$ in Fig.~\ref{Fig:2}D (using standard SiO$_2$ cladding), the EDWL would achieve a uniform power distribution across the C- and L-bands.
While erbium ions emit more strongly in the C-band, the reduced propagation losses and higher saturation power at longer wavelengths balance the output power.
Thus, minimizing cavity loss is crucial to maximizing lasing power and achieving spectral uniformity across the Er emission bandwidth.

Figure \ref{Fig:3}E presents the on-chip laser power as a function of on-chip pump power, revealing a lasing threshold of 17.5~mW at 1480~nm pumping and a slope efficiency of approximately 24$\%$, which can be further optimized by reducing the coupling loss and the cavity loss.

\begin{figure*}[t!]
\centering
\includegraphics[width=\textwidth]{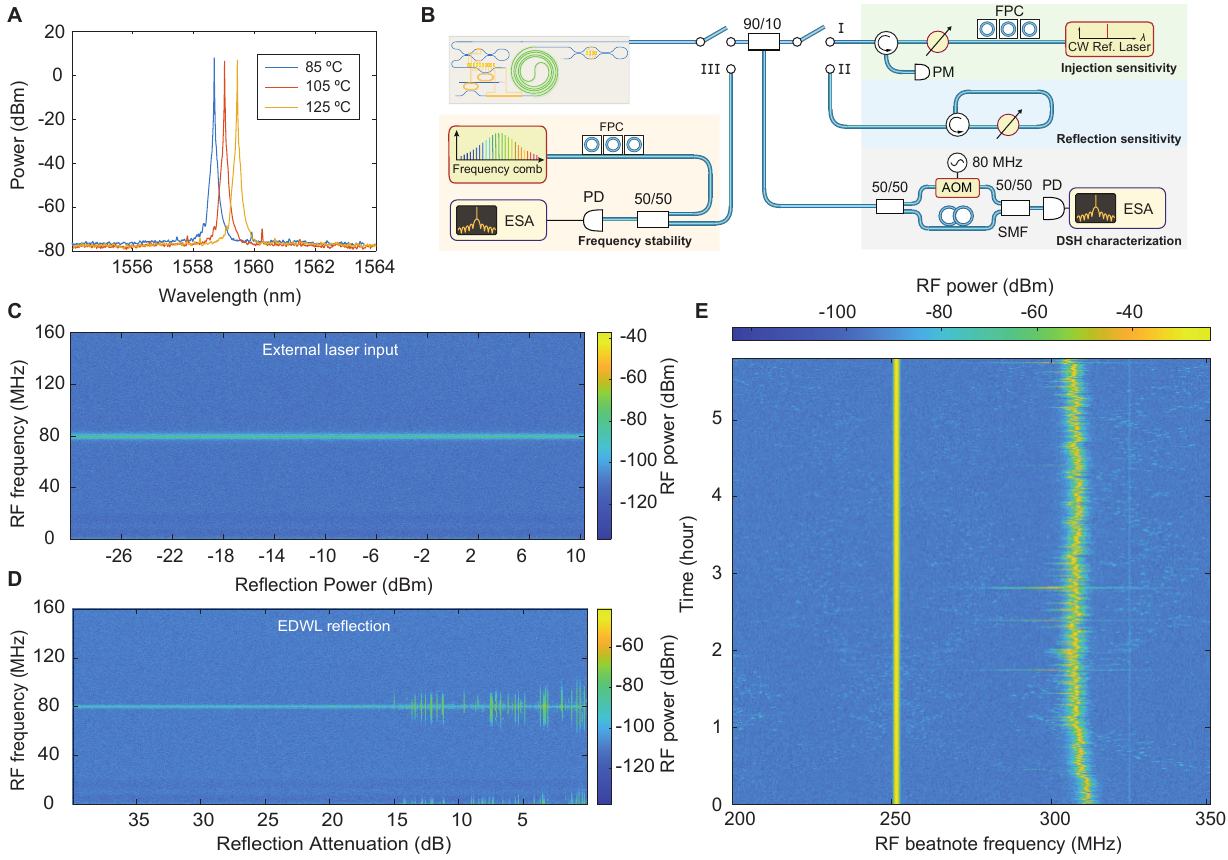}
\caption{
\footnotesize
\textbf{Stability characterization of the Er-based integrated tunable laser.}
(\textbf{A}) Optical spectra of lasing at device temperatures from 85~$^\circ\text{C}$ to 125~$^\circ\text{C}$, exceeding the temperature limit of most commercial III-V semiconductor lasers.
(\textbf{B}) Experimental setup for laser stability characterization, evaluating insensitivity to external reflections (Path I), self-reflections (Path II), and lasing wavelength drift through heterodyne beatnote measurement with a stable frequency comb (Path III).
The RF beatnote is measured by a delayed self-heterodyne (DSH) interferometric measurement. 
The center frequency of 80 MHz is determined by the RF frequency applied to the acousto-optic modulator (AOM). SMF: single mode fiber delay line, FPC: fiber polarization controller.
(\textbf{C}) Measured spectrogram of the laser DSH beatnote when increasing the power of an injected external laser at the same wavelength, within a detuning of 350 MHz.
(\textbf{D}) Measured spectrogram of the laser DSH beatnote under increasing self-reflection levels.
The optical path with circulator was modified to reflect the laser through a variable optical attenuator (VOA).
(\textbf{E}) Measured spectrogram of the laser heterodyne beatnote with an optical reference system (FC1500, Menlo Systems) over 6 hours, using remote pumping to mitigate thermal instabilities from direct integration.
The peak at 250 MHz corresponds to the frequency comb repetition rate.
The peak at 310 MHz is a beat of a comb line with the laser, drifting less than 15 MHz over 6 hours.}
\label{Fig:4}
\end{figure*}
We evaluated the single-sided power spectral density (PSD) of frequency noise by processing the in-phase and quadrature components of the sampled beatnote between the EDWL and an external cavity diode laser (ECDL, Toptica CTL) using Welch's method~\cite{welch1967use} (Fig. \ref{Fig:3}F).
For a free-running EDWL lasing at 1592~nm (blue trace, EDWL 1), the frequency noise PSD beyond the offset frequency of 1~MHz exhibits a plateau of $\mathrm{h_0}=24.6~\mathrm{Hz}^2/\mathrm{Hz}$, corresponding to a Lorentzian linewidth $\pi \mathrm{h_0}$ of 77.3~Hz.
A full width at half-maximum (FWHM) linewidth of 99.7~kHz, associated with Gaussian contributions, was obtained by integrating the frequency noise from 1/T$_0$ up to the frequency where frequency noise PSD $S_\mathrm{v}(f)$ intersects the $\beta$-separation line $8 \ln (2) \cdot f/ \pi^2$ (dashed grey) at T$_0=$1~ms measurement time.
A relaxation oscillation peak is observed at 5~MHz, calculated as $f_\mathrm{r} \approx \frac{1}{2\pi}\sqrt{\frac{\mathrm{P_{cav}}\kappa}{\mathrm{P_{sat}}\tau}}$, where $\mathrm{P_{sat}}$ is the saturation power of the gain medium, $\mathrm{P_{cav}}$ is the laser cavity power, $\kappa$ is the cold cavity loss rate, and $\tau$ is the erbium ion upper-state lifetime.
In a fully packaged laser, represented by the red trace (EDWL 2) measured at 1588 nm, a lower FWHM linewidth of 90.4~kHz is achieved by integrating over a 1~ms measurement time, attributed to reduced contributions from mechanical vibrations at lower offset frequencies.
The relaxation oscillation frequency in EDWL 2 occurs at a lower offset of 0.5~MHz, primarily due to differences in laser cavity power.
The frequency noise PSD of the reference ECDL (in grey) was obtained via heterodyne measurement with a commercial stabilized fiber-based laser (Koheras ADJUSTIK) near 1550~nm.
A similar PSD plateau $\mathrm{h_0}$ is observed beyond 1~MHz, aligning with the characteristics of EDWL 1 and EDWL 2.
Noise features below 10~kHz offsets are also comparable, indicating dominant contributions from the ECDL in this range for the measured EDWL PSDs.
Using a reference laser with lower frequency noise beyond 1~MHz and below 10~kHz could further reduce the resultant Lorentzian linewidth $\uppi \mathrm{h_0}$ and FWHM linewidth of the EDWLs, demonstrating a performance comparable to or surpassing that of the ECDL in these frequency ranges. 

Among the various factors influencing the EDWL's frequency stability, thermo-refractive index noise (TRN) emerges as a significant contributor at offset frequencies below 1~MHz.
Using established models \cite{kondratiev2018thermorefractive,huang2019thermorefractive}, we computed the TRN for a single ring resonator in the Vernier filter and estimated its contribution to the overall laser cavity (detailed in Supplementary Note 6).
The simulated TRN is shown as a dashed red line in Fig. \ref{Fig:3}F.
Additionally, intensity noise from the 1480~nm pump laser contributes to the EDWL frequency noise through cavity and Vernier filter ring heating \cite{noisetrans}.
To quantify this, we modulated the pump laser’s drive current with a weak sinusoidal signal, measuring the resulting RIN on an electrical spectrum analyzer (ESA) and the corresponding peak in the laser frequency noise spectrum \cite{lucas2020ultralow}.
The transduction function, calculated as the ratio of these peaks, was used to scale the pump RIN in Fig. \ref{Fig:3}G (Supplementary Note 5), estimating the noise contribution from the pump laser diode (violet trace in Fig. \ref{Fig:3}F).
This contribution dominates at offset frequencies above 1~MHz.
A comparison of EDWL frequency noise obtained with varied pump RIN is provided in Supplementary Note 7.

The RIN of the EDWL was measured in Fig. \ref{Fig:3}G via photodetection using a DC-coupled photodetector (Thorlabs DET08CFC) with 0.9 A/W responsivity at 1550 nm.
The measured EDWL RIN (blue) reaches values as low as $-150~\mathrm{dBc/Hz}$ at mid-range offset frequencies from 5 kHz to 1 MHz.
The EDWL RIN closely follows the trend of the pump diode (purple), indicating a strong correlation between pump intensity noise and the EDWL's RIN performance.
The observed decrease of RIN beyond 1~MHz offset frequency is due to the drop-port filtering of the Vernier filter.
Further improvements in pump diode noise could enhance the EDWL's overall noise performance.

\subsection*{Laser emission robustness against temperature and reflection}

Semiconductor lasers prevail in industrial applications due to their compactness and maturity in mass production.
They are highly sensitive to elevated temperatures, which lead to increased carrier recombination, higher threshold currents, reduced efficiency, and drifting of the gain center~\cite{levinshtein1997handbook,agrawal2013semiconductor,chen1983carrier}, except with improved thermal management and laser design~\cite{sun2024high}.
In contrast, the erbium-doped gain medium is inherently temperature-insensitive due to its narrow, atomic-like optical transitions shielded by filled 5s and 5p electron shells, which suppress phonon interactions and stabilize erbium energy levels against temperature fluctuations~\cite{desurvire1995erbium}.
As a result, erbium-doped gain media exhibit minimal change in their emission cross-section and population inversion efficiency with temperature.
This ensures that erbium-doped lasers maintain stable output power and linewidth, even at elevated temperatures, without the performance degradation commonly seen in semiconductor-based lasers.

Figure \ref{Fig:4}A presents the laser emission spectrum at 85 $^\circ\text{C}$ (6.3 mW), 105 $^\circ\text{C}$ (4.4 mW), and 125 $^\circ\text{C}$ (5 mW), demonstrating stable operation with a slight power reduction compared to room temperature (10 mW), primarily due to coupling drifts of the coupling fibers.

Another significant challenge faced by integrated lasers is back-reflections, which degrade coherence, introduce frequency noise, and cause intolerable nonlinearities for many analog and digital applications \cite{tkach1986regimes}.
We characterized the stability of the Er-doped integrated laser under varying reflection conditions in Figure \ref{Fig:4}.
To evaluate the effect of reflections, we used the experimental setup shown in Fig \ref{Fig:4}B, where reflection sensitivity was characterized by injecting controlled back-reflections into the laser cavity.
Figures \ref{Fig:4}C and \ref{Fig:4}D present the spectrograms of the laser beatnote with external and self-reflections, respectively, monitored via delayed self-heterodyne detection.
The RF beatnote, centered at 80 MHz, corresponds to the RF frequency applied to the acousto-optic modulator (AOM).
The feedback levels were systematically varied from -30 dBm to +10 dBm for external reflections.
The laser demonstrated resilience to these reflections, maintaining a consistent frequency response without coherence collapse across the tested reflection levels.
Notably, the laser maintained stability with self-reflections up to 14 dB attenuation, on par with, or better than conventional ECDLs such as the Toptica DL pro, which typically require >30 dB optical isolation to mitigate back-reflections.
This robustness against back-reflections, attributed to the microresonator drop ports functioning as intrinsic bandpass filters in the EDWL, suggests potential isolator-free operation, simplifying photonic designs and enhancing scalability in demanding environments.

Next we characterized the long-term frequency stability of the integrated Er-doped Si$_3$N$_4$ laser (Fig \ref{Fig:4}E) by monitoring the RF beatnote spectrum from its heterodyne interference with a fully-stabilized optical frequency comb (FC1500, Menlo Systems).
The Er integrated laser is mounted in a 14-pin butterfly package with UHNA-7 fibers for Er pumping (Fig \ref{Fig:3}B), while the III-V 1480 pump laser diode is positioned remotely to isolate heat-induced instability.
The beatnote with center frequency of 250 MHz, derived from two comb lines of the optical frequency comb source, was used to monitor and benchmark stability.
The laser-comb line beatnote was tracked over a six-hour period.
The spectrum in Fig. \ref{Fig:4}E shows a frequency drift within 15 MHz, demonstrating strong stability in our Er:Si$_3$N$_4$ laser.
This stability, attributed to the monolithic design with a high-$Q$ cavity and efficient thermal management, meets the stringent demands of advanced applications such as precision sensing, LiDAR, and high-coherence optical communications, where narrow linewidth, high stability, and disturbance resistance are critical.
 
\subsection*{Summary}
In summary, we demonstrated the first Er-doped waveguide laser fabricated via wafer-scale processes, achieving a 75 dB SMSR and quasi-full C- and L-band tunability.
Using 200-nm thick Si$_3$N$_4$ photonic integrated circuits, the implantation energy was reduced to below 500~keV, with scalability demonstrated through wafer-scale implantations in a commercial 300~mm tool.
The laser delivers up to 36~mW fiber-coupled output power with a Lorentzian linewidth below 95~Hz and a frequency drift within 15~MHz over 6 hours.
A key advantage of the Er:Si$_3$N$_4$ Vernier laser architecture is the temperature insensitivity, enabling operation at up to 125~$^\circ\text{C}$ without significant performance degradation.
The architecture also supports remote pumping, ensuring environmental stability and functionality in harsh conditions without requiring costly sealed packaging.
Future improvements in gain coefficient and cavity loss could achieve fiber laser-level coherence, establishing the laser as a compact, mass-producible platform for next-generation optical systems, enabling applications in coherent sensing, LiDAR, analog optical links, and coherent communications.

\medskip
\subsection*{Acknowledgments}

\medskip
\begin{footnotesize}
\noindent \textbf{Funding Information}: This work was supported by the Swiss National Science Foundation under grant no. 216493 (HEROIC).

\noindent \textbf{Acknowledgments}: 
We thank Professor Carsten Ronning for the Er photoluminescence measurements. The Si$_3$N$_4$ samples were fabricated in the EPFL Center of MicroNanoTechnology (CMi).

\noindent \textbf{Author contributions}: 
Y.L. conceived the idea and concept.
X.J. and Y.L. designed Si$_3$N$_4$ waveguide laser chips.
X.J. and Z.Q. fabricated the Si$_3$N$_4$ samples. 
T.K. and J.C.O. performed the ion implantation on the Si$_3$N$_4$ samples.
X.J, X.Y., Y.L. performed the experiments with support from G.L., S.B and Z.Q.
X.J, Y.L., X.Y. carried out data analysis and simulations. 
A.V. designed and performed the device packaging.
Y.L. and X.J. wrote the manuscript with input from all co-authors.
T.J.K. and Y.L. supervised the project.

\noindent \textbf{Data Availability}:
The data used to produce the plots within this work will be released on the repository \texttt{Zenodo} upon publication of the work. 

\noindent \textbf{Code Availability}: 
The data used to produce the plots within this work will be released on the repository \texttt{Zenodo} upon publication of the work.

\noindent \textbf{Competing interests}
T.J.K. is a cofounder and shareholder of LiGenTec SA, a start-up company offering Si$_3$N$_4$ photonic integrated circuits as a foundry service.
T.J.K. is a cofounder and shareholder of EDWATEC SA, a start-up company offering optical amplifiers on chip.


\end{footnotesize}


\pretolerance=0
\bigskip
\bibliographystyle{apsrev4-2}
\bibliography{zotero_updated_v1}

\section*{Methods}
\subsection*{Photonic integrated circuit fabrication}
For the fabrication of passive, ultra-low loss \ce{Si3N4} photonic integrated circuits, a single-crystalline 100 mm Si wafer was wet-oxidized with $8.0~\mathrm{\upmu m}$ SiO$_2$ to provide a bottom cladding which can effectively isolate light confined in the thin \ce{Si3N4} waveguide from the silicon substrate, including at edge couplers.
The gain-section waveguides, with a 5$\times$0.2~$\upmu$m$^2$ cross-section and a 3.02~$\upmu$m$^2$ mode area (Fig. \ref{Fig:1}G), are formed from 200~nm thick low-pressure chemical vapor deposition (LPCVD) Si$_3$N$_4$ films.
The LPCVD Si$_3$N$_4$ films exhibit an RMS roughness of 0.3~nm and uniformity of $\pm0.6\%$ across a 4-inch wafer.
Following deposition, Si$_3$N$_4$ films are annealed at 1200$^\circ$C for 11 hours to eliminate excess H$_2$ and break N-H and Si-H bonds, which exhibit absorption in the 1500–1538~nm wavelength range \cite{germann2000silicon,bauters2011planar}.
This annealing process results in a 3.5$\%$ thickness reduction in the 200~nm Si$_3$N$_4$ layer due to film densification and hydrogen effusion \cite{jiang2016effect}.
In contrast to post-waveguide annealing in thick Si$_3$N$_4$ films~\cite{ji2024efficient}, immediate annealing after LPCVD ensures precise thickness control and preserves the waveguide shape, preventing deformation caused by tensile stress in low-aspect-ratio waveguides (Supplementary Note 2).
The waveguide patterns are defined using deep ultraviolet (DUV) stepper lithography (KrF 248~nm) with a resolution of 180~nm.
Si$_3$N$_4$ waveguides are formed via anisotropic dry etching with CHF$_3$ and SF$_6$ gases, yielding vertical, clean, and smooth sidewalls while minimizing polymer redeposition from etching byproducts.

After ion implantation, we annealed the chip at \SI{1000}{\degreeCelsius} for 1 hour in \ce{N2} to optically activate the erbium ions and restore the implantation-induced defects.
A 3~$\mathrm{\upmu}$m-thick hydrogen-free, low-loss \ce{SiO2} cladding was subsequently grown at \SI{300}{\degreeCelsius} using inductively coupled plasma-enhanced chemical vapor deposition with \ce{SiCl4} and \ce{O2} as precursors~\cite{qiu_hydrogen-free_2023}.
Integrated Pt/Ti microheaters were fabricated on top of the \ce{SiO2} upper claddings to enable the thermo-optic tuning of microresonator resonance frequency and phase shift.
After a fully wafer-scale fabrication process, the wafer was separated into dies of individual chip-scale lasers (Fig.~\ref{Fig:1}D) via deep etching of \ce{SiO2} and subsequent deep reactive ion etching (DRIE) of Si using the Bosch method, followed by backside grinding.

\end{document}


\title{Supplementary Information for: Full C- and L-band tunable erbium-doped integrated lasers via scalable manufacturing}

\author{Xinru Ji$^{1,2}$, 
				Xuan Yang$^{1,2}$, 
				Yang Liu$^{1,2}$,
			    Zheru Qiu$^{1,2}$, 
			    Grigory Lihachev$^{1,2}$, 
			    Simone Bianconi$^{1,2}$,
			    Jiale Sun$^{1,2}$, 
                Andrey Voloshin$^{1,2}$,
                Taegon Kim$^{3}$,
                Joseph C. Olson$^{3}$,
				and Tobias J. Kippenberg$^{1,2\dag}$}
\affiliation{
$^1$Institute of Physics, Swiss Federal Institute of Technology Lausanne (EPFL), CH-1015 Lausanne, Switzerland\\
$^2$Center for Quantum Science and Engineering, Swiss Federal Institute of Technology Lausanne (EPFL), CH-1015 Lausanne, Switzerland\\
$^3$Varian Semiconductor, Applied Materials, Gloucester, MA 01930, United States
}

\setcounter{equation}{0}
\setcounter{figure}{0}
\setcounter{table}{0}

\setcounter{subsection}{0}
\setcounter{section}{0}
\setcounter{secnumdepth}{3}

 \begin{abstract}
Supplementary Information for this manuscript includes a comparative study of ion implantation in high- and low-confinement Si$_3$N$_4$ waveguides, theoretical analysis of Er integrated laser performance, detailed sample fabrication process, broadband tunable loop mirror design and characterization, experimental methods for laser RIN and linewidth measurements, analysis of laser frequency noise dependence on pump laser characteristics, and laser frequency noise originating from heater resistivity drift.
 \end{abstract}

\maketitle
{\hypersetup{linkcolor=blue}\tableofcontents}
\newpage


\section{Ion implantation parameters for Er-doped Si$_3$N$_4$ devices: thick vs. thin waveguides}

\begin{table}[h!]
\centering
\caption{Comparison of fabrication requirements and ion implantation parameters for thick (700 nm) and thin (200 nm) Er-doped Si$_3$N$_4$ devices}
\setlength{\tabcolsep}{4pt}
\setlength{\extrarowheight}{7pt}
\begin{tabular}{>{\centering\arraybackslash}p{2cm}|>{\centering\arraybackslash}p{2cm}|>{\centering\arraybackslash}p{2cm}|>{\centering\arraybackslash}p{1.2cm}|>{\centering\arraybackslash}p{2cm}|>{\centering\arraybackslash}p{2.5cm}|>{\centering\arraybackslash}p{2.2cm}|>{\centering\arraybackslash}p{1.5cm}}
\hline 
\textbf{Si$_3$N$_4$ waveguide thickness (nm)} & \textbf{Fabrication steps for passive waveguides} & \textbf{Implantation machine model} & \textbf{Ion energy (keV)} & \textbf{Ion fluence (cm$^{-2}$)} & \textbf{Beam current density (cm$^{-2}$)} & \textbf{Implantation area (cm$^{2}$)}& \textbf{Net time}\\
\hline 
\multirow{3}{*}{200} & \multirow{3}{*}{4$^{*}$} & \multirow{3}{*}{VIISta HE} & 480 & 3.20$\times$10$^{15}$ & 0.15 $\mu$A &\multirow{2}{*}{$\sim$707}& 60 min. \\
& & & 270 & 1.50 $\times$10$^{15}$ & 0.15 $\mu$A & (12$^{\prime\prime}$ wafer) & 30 min. \\
& & & 130 & 1.10 $\times$10$^{15}$ & 0.3 $\mu$A & & 10 min. \\
\hline 
\multirow{3}{*}{700} & \multirow{3}{*}{6$^{**}$} & \multirow{3}{*}{Tandem} & 2000 & 4.50$\times$10$^{15}$ & 13.8 - 304.5 pA &\multirow{3}{*}{$\sim$0.72} & 18 hours \\
& & & 1416 & 3.17$\times$10$^{15}$ & 41.5 - 152.2 pA & & 19 hours \\
& & & 955 & 2.34$\times$10$^{15}$ & 96.9 - 179.9 pA & & 9.5 hours \\
\hline
\end{tabular}
 \vspace{1mm} 

    \parbox[b]{0.9\linewidth}{
    \footnotesize
    $^*$ 4 steps: Deposition, patterning, etching, and annealing of the Si$_3$N$_4$ waveguides~\cite{ji2024efficient}. \\
    $^{**}$ 6 steps: Patterning, preform etching, preform reflow, deposition, planarization, and annealing of the Si$_3$N$_4$ waveguides~\cite{liu2021high}.
    }
\label{SI:tab:1}
\end{table}

In the development of erbium (Er) doped silicon nitride (Si$_3$N$_4$) waveguide devices, the choice of ion implantation parameters—particularly ion energy and dose—plays a critical role in determining the feasibility and scalability of the fabrication process.
In our prior works \cite{liu_photonic_2022,liu_fully_2023}, thick silicon nitride waveguides ($\sim$700 nm) with high confinement required high-energy ion implantations up to 2 MeV to achieve optimal optical mode overlap with the erbium ions.
However, such high-energy implantations are not commonly used in industrial applications due to their high cost and limited accessibility, especially in large-scale processes.
Additionally, high-energy implants typically require advanced equipment setups like gas-insulated electrostatic accelerators, which further increase operational complexity.
To address these challenges, there is a shift towards using thinner silicon nitride layers ($\sim$200 nm), allowing for ion implantation at significantly lower energies, around 480 keV.
This reduction in ion energy aligns more closely with the implantation parameters used in current microelectronics manufacturing, where standard implanters typically operate below 600 keV for most high-dose applications \cite{rubin2003ion}.

\begin{figure*}[t!]
\centering
\includegraphics[width=0.8\textwidth]{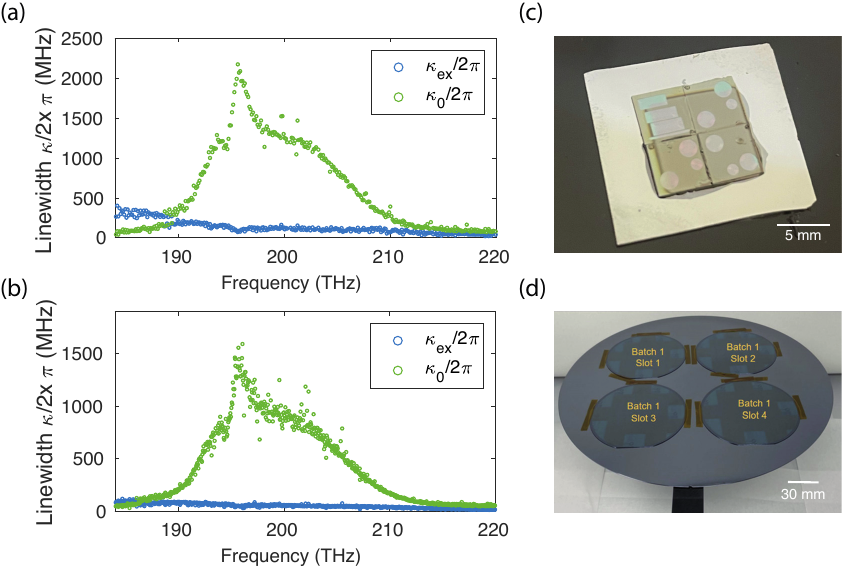}
\caption{
\footnotesize
\textbf{Comparison of high-energy and low-energy Er implantations}
(\textbf{a}) Measured optical losses (intrinsic linewidths $\kappa_0/2 \pi$) of the Er-implanted Si$_3$N$_4$ microring resonator with 2 MeV Er implantation energy.
The implantation parameters are given in SI table \ref{SI:tab:1}.
The microring resonator features a cross-section of 2.1$\times$0.7 $\mu$m$^2$ and a free spectral range (FSR) of 100 GHz.
The intrinsic linewidths present a characteristic Er absorption profile.
(\textbf{b}) Measured optical losses (intrinsic linewidths $\kappa_0/2 \pi$) of the Er-implanted Si$_3$N$_4$ microring resonator with 480 keV Er implantation energy, using the implantation parameters given in SI table \ref{SI:tab:1}.
The Si$_3$N$_4$ waveguide has a cross-section of 5$\times$0.2 $\mu$m$^2$ and an FSR of 50 GHz.
(\textbf{c}) Picture of the high-energy implanted samples characterized in (\textbf{a}).
The total implantation area is 8.5$\times$8.5 mm$^2$.
(\textbf{d}) Picture of the low-energy implanted samples characterized in (\textbf{b}).
This picture shows four 4-inch wafers mounted on a 12-inch wafer during implantation.
The total implantation area is up to 707 cm$^2$, corresponding to a 12$^{\prime\prime}$ wafer.
}
\label{Fig:Imp_comp}
\end{figure*}

Supplementary table~\ref{SI:tab:1} summarizes key Er ion implantation parameters from our recent runs, including fabrication steps for passive waveguide preparation, ion energy, fluence, beam current density, and net implantation time, which generally scales with ion fluence.
The 200 nm waveguide height is carefully chosen to optimize the overlap with Er ions while aligning with available implantation energies in the semiconductor industry.

A comparison of Er absorption profiles, characterized by intrinsic linewidths in Er:Si$_3$N$_4$ microring resonators, is provided in Fig. \ref{Fig:Imp_comp}(a)(b).
Fig. \ref{Fig:Imp_comp}(a) shows the intrinsic linewidth ($\kappa_0/2\pi$) of a 2 MeV implanted resonator with a 2.1$\times$0.7 $\mu$m$^2$ cross-section and 100 GHz FSR, while Fig. \ref{Fig:Imp_comp}(b) depicts the intrinsic linewidth of a 480 keV implanted resonator with a 5$\times$0.2 $\mu$m$^2$ cross-section and 50 GHz FSR.
Implantation parameters for both resonators are detailed in Supplementary Table~\ref{SI:tab:1}.
$\kappa_0/2\pi$, measured via frequency-comb-assisted broadband laser spectroscopy, reveal characteristic Er absorption profiles in both high-energy and low-energy samples.
Differences in peak absorption (maximum of $\kappa_0/2\pi$) arise from variations in Er profiles optimized for overlap factors.
By scaling the Er ion implantation dose in Fig. \ref{Fig:Imp_comp}(b) to match the peak absorption in Fig. \ref{Fig:Imp_comp}(a), similar peak absorption can be achieved in the low-energy implanted sample.
However, in thinner Si$_3$N$_4$ waveguides, closer Er ion spacing compared to thicker waveguides increases the likelihood of ion clustering at high implantation doses. Elevated Er concentrations further induce pair-induced quenching, caused by energy transfer between closely spaced ions, which hinders full population inversion and reduces quantum efficiency \cite{myslinski1997effects,dong2004concentration}.
To mitigate these effects while maintaining high gain and efficiency, implantation parameters for 200 nm Si$_3$N$_4$ waveguides are carefully optimized.

By moving towards lower-energy implantations, we aim to achieve greater availability and economic viability, enabling wafer-scale doping of Si$_3$N$_4$ waveguides (up to 12$^{\prime\prime}$ wafers with the VIISta HE implanter, Fig. \ref{Fig:Imp_comp}(d)), compared to the previous high-energy approach, which was typically limited to small-area doping ($\sim$8.5$\times$8.5 mm$^2$ with the Tandem implanter, Fig. \ref{Fig:Imp_comp}(c)).
The VIISta HE implanter provides a dose uniformity within 0.5$\%$ (1 $\sigma$) and an angle accuracy of $\pm$0.1$^{\circ}$ for 8-inch wafers.
The shift to thinner nitride films and lower-energy implantations not only improves cost-efficiency but also enhances device throughput by leveraging high-current and medium-current implanters already prevalent in the semiconductor industry.
Aligning with these established protocols makes the fabrication of Er-doped Si$_3$N$_4$ devices more scalable and commercially viable for practical applications.

\section{Impact of annealing and etching sequence on Si$_3$N$_4$ waveguides}

\begin{figure*}[h!]
\centering
\includegraphics[width=\textwidth]{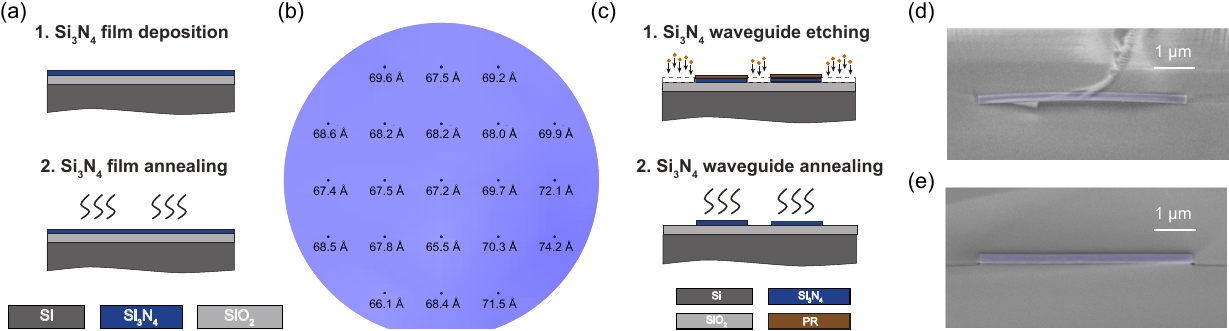}
\caption{
\footnotesize
\textbf{Effect of annealing-etching order on Si$_3$N$_4$ waveguide properties.} 
(\textbf{a}) Pre-fabrication of Si$_3$N$_4$ waveguides: Low-pressure chemical vapor deposition of Si$_3$N$_4$ thin films and film annealing at 1200$^\circ$C for 11 hours.
(\textbf{b}) Shrinkage of a 200~nm Si$_3$N$_4$ film thickness after annealing.
(\textbf{c}) Schematic of Si$_3$N$_4$ waveguide fabrication process, showing etching and annealing steps.
(\textbf{d}) The scanning electron microscopic image of Si$_3$N$_4$ waveguide fabricated with etching-annealing sequence.
(\textbf{e}) The scanning electron microscopic image of Si$_3$N$_4$ waveguide fabricated with annealing-etching sequence.
}
\label{Fig:anneal}
\end{figure*}

We investigate the impact of the etching-annealing sequence on passive Si$_3$N$_4$ waveguides.

In the fabrication process for the EDWL devices discussed in the main manuscript, 200 nm LPCVD Si$_3$N$_4$ is deposited on a silicon substrate with an 8 $\mu$m wet oxide (WOX) layer.
Si$_3$N$_4$ waveguides are fabricated through deep ultraviolet (DUV) stepper lithography, followed by dry etching and a high-temperature annealing.
Depending on the conditions, high-temperature annealing generally induces film shrinkage in LPCVD Si$_3$N$_4$ films~\cite{jiang2016effect}. Here, we anneal the Si$_3$N$_4$ waveguide at 1200 $^\circ$C for 11 hours, necessitating an investigation of the resulting thickness reduction.

We measured a 6.8 nm average thickness reduction in a 200 nm Si$_3$N$_4$ film after annealing (Supplementary Figure~\ref{Fig:anneal}(b)).
Such shrinkage can significantly affect devices sensitive to waveguide height variations.
For instance, in waveguide Bragg gratings, a thickness variation of 7 nm can cause a central reflection frequency shift on the order of terahertz.
Additionally, due to the large intrinsic tensile stress of LPCVD Si$_3$N$_4$ films, the 5 $\mu$m $\times$ 200 nm waveguides tend to bend unpredictably after annealing, as shown in the SEM image in Supplementary Figure~\ref{Fig:anneal}(d).
This bending affects the optical mode distribution, leading to light scattering, which may impact the performance of the erbium-doped gain section in EDWL devices. 

However, reversing the sequence of Si$_3$N$_4$ waveguide etching and annealing effectively mitigates this bending.
Supplementary Figure~\ref{Fig:anneal}(e) presents the cross-section of a Si$_3$N$_4$ waveguide fabricated with the annealing-etching sequence, where no bending is observed, preserving a rectangular cross-sectional shape.
The annealing-etching sequence also allows improved control of waveguide thickness, as it facilitates thickness measurement at the film level.

\section{Broadband tunable loop mirror design and characterization}

The tunable broadband loop mirror used in the EDWL in the main text, illustrated in Supplementary figure~\ref{Fig:TLM}, is based on a looped Mach-Zehnder Interferometer (MZI) structure.
The device features two input fields, $E_\mathrm{i 1}$ and $E_{\mathrm{i 2}}$, and two output fields, $E_{\mathrm{o 1}}$ and $E_{\mathrm{o 2}}$.
Directional couplers within the MZI, defined by coupling ratios $k_{1}$ and $k_{2}$, govern the power distribution between the two arms.
Phase tuning is implemented using integrated metal heaters, which introduce controlled phase shifts to dynamically adjust the transmission and reflection characteristics of the loop mirror, enabling broadband tuning of the output fields $E_{\mathrm{o 1}}$ and $E_{\mathrm{o 2}}$.

The mathematical relationship between $E_{\mathrm{o 1}}$, $E_{\mathrm{o 2}}$ and $E_{\mathrm{i 1}}$, $E_{\mathrm{i 2}}$ writes:

\begin{equation}
\begin{aligned}
\left[\begin{array}{l}
E_\mathrm{o1} \\
E_\mathrm{o2}
\end{array}\right] &= 
\left[\begin{array}{cc}
\sqrt{1-k_1} & -i \sqrt{k_1} \\
-i \sqrt{k_1} & \sqrt{1-k_1}
\end{array}\right] \times
\left[\begin{array}{cc}
\exp \left(-i \beta L_1 - i \Delta \varphi\right) & 0 \\
0 & \exp \left(-i \beta L_2\right)
\end{array}\right]\\
&\quad \times \left[\begin{array}{cc}
\sqrt{1-k_2} & -i \sqrt{k_2} \\
-i \sqrt{k_2} & \sqrt{1-k_2}
\end{array}\right]  \times
\left[\begin{array}{cc}
\exp (-i \beta L) & 0 \\
0 & \exp (-i \beta L)
\end{array}\right] \\
&\quad \times \text{ swap } \times
\left[\begin{array}{cc}
\sqrt{1-k_2} & -i \sqrt{k_2} \\
-i \sqrt{k_2} & \sqrt{1-k_2}
\end{array}\right] \times
\left[\begin{array}{cc}
\exp \left(-i \beta L_1 - i \Delta \varphi\right) & 0 \\
0 & \exp \left(-i \beta L_2\right)
\end{array}\right] \\
&\quad \times \left[\begin{array}{cc}
\sqrt{1-k_1} & -i \sqrt{k_1} \\
-i \sqrt{k_1} & \sqrt{1-k_1}
\end{array}\right] \times
\left[\begin{array}{l}
E_\mathrm{i1} \\
E_\mathrm{i2}
\end{array}\right]
\end{aligned}
\label{eq:1}
\end{equation}

\begin{figure*}[t!]
    \centering
    \includegraphics[width=0.9\textwidth]{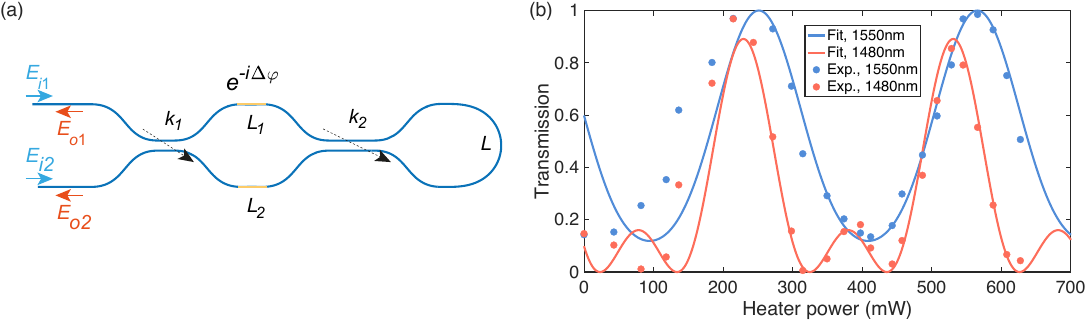}
    \caption{
    \footnotesize
    \textbf{Broadband tunable loop mirror characteristics.}
    (\textbf{a}) Schematic of the tunable loop mirror. $E_\mathrm{i1}$ and $E_\mathrm{i2}$: input fields; $E_\mathrm{o1}$ and $E_\mathrm{o2}$: output fields; $k_1$ and $k_2$: power coupling ratios in the directional couplers; $L_1$ and $L_2$: lengths of the Mach-Zehnder Interferometer (MZI) arms.
    (\textbf{b}) Simulated and experimental dependence of transmitted output field on applied heater power in the MZI arm.
     Simulation parameters for 1550 nm: $E_\mathrm{i1}$=0, $E_\mathrm{i2}$=1, $k_1=k_2$=0.09; for 1480 nm: $E_\mathrm{i1}$=0, $E_\mathrm{i2}$=1, $k_1$=0.16, $k_2$=0.3.
    }
    \label{Fig:TLM}
    \end{figure*}
    
\begin{figure*}[b!]
    \centering
    \includegraphics[width=0.9\textwidth]{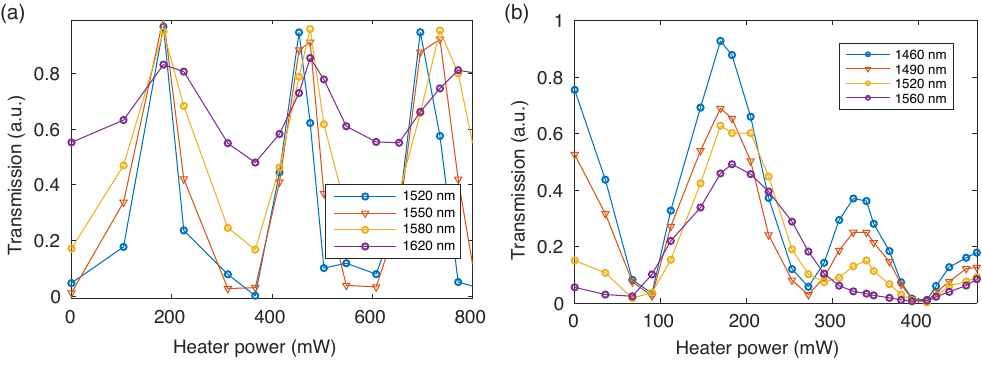}
    \caption{
    \footnotesize
    \textbf{Heater power-dependent tunable loop mirror characterization.}
    Transmission of (\textbf{a}) the broadband reflector and (\textbf{b}) mode-selective reflector in the EDWL.
    }
    \label{Fig:TLM2}
    \end{figure*}
    
The output fields $E_{\mathrm{o 1}}$, $E_{\mathrm{o 2}}$ are computed as:

\begin{equation}
\begin{aligned}
E_{o1} = & \; \left(-i e^{-i \Delta \varphi} \sqrt{1-k_1} \sqrt{k_2} - i \sqrt{k_1} \sqrt{1-k_2}\right) \\
& \quad \times \left[E_0 \left(-\sqrt{k_1} \sqrt{k_2} + e^{-i \Delta \varphi} \sqrt{1-k_1} \sqrt{1-k_2}\right) + E_1 \left(-i e^{-i \Delta \varphi} \sqrt{k_1} \sqrt{1-k_2} - i \sqrt{1-k_1} \sqrt{k_2}\right)\right] \\
& + \left(-\sqrt{k_1} \sqrt{k_2} + e^{-i \Delta \varphi} \sqrt{1-k_1} \sqrt{1-k_2}\right) \\
& \quad \times \left[E_0 \left(-i e^{-i \Delta \varphi} \sqrt{1-k_1} \sqrt{k_2} - i \sqrt{k_1} \sqrt{1-k_2}\right) + E_1 \left(\sqrt{1-k_1} \sqrt{1-k_2} - e^{-i \Delta \varphi} \sqrt{k_1} \sqrt{k_2}\right)\right] \\
E_{o2} = & \; \left(-i e^{-i \Delta \varphi} \sqrt{k_1} \sqrt{1-k_2} - i \sqrt{1-k_1} \sqrt{k_2}\right) \\
& \quad \times \left[E_0 \left(-i e^{-i \Delta \varphi} \sqrt{1-k_1} \sqrt{k_2} - i \sqrt{k_1} \sqrt{1-k_2}\right) + E_1 \left(\sqrt{1-k_1} \sqrt{1-k_2} - e^{-i \Delta \varphi} \sqrt{k_1} \sqrt{k_2}\right)\right] \\
& + \left(\sqrt{1-k_1} \sqrt{1-k_2} - e^{-i \Delta \varphi} \sqrt{k_1} \sqrt{k_2}\right) \\
& \quad \times \left[E_0 \left(-\sqrt{k_1} \sqrt{k_2} + e^{-i \Delta \varphi} \sqrt{1-k_1} \sqrt{1-k_2}\right) + E_1 \left(-i e^{-i \Delta \varphi} \sqrt{k_1} \sqrt{1-k_2} - i \sqrt{1-k_1} \sqrt{k_2}\right)\right]
\end{aligned}
\label{eq:2}
\end{equation}

In Supplementary equations \ref{eq:1} and \ref{eq:2}, $\beta$ is the propagation constant, and $e^{-i \Delta \varphi}$ denotes the additional phase shift introduced by micro-heaters.
The nonlinear phase shift over the closed loop of length $L$ is neglected, and the MZI arm lengths $L_1$ and $L_2$ are set equal.

Supplementary Figure \ref{Fig:TLM}(b) presents experimental and simulated transmission characteristics as a function of applied heater power in a tunable loop mirror for 1480 nm and 1550 nm inputs.
In these measurements, the input field is injected at one port ($E_\mathrm{i1}$=0 and $E_\mathrm{i2}$=1), and transmission is defined as the output at the opposite port.
For both wavelengths, transmission is tunable from 0 to 1, demonstrating the loop mirror’s ability to switch between fully transmissive and fully reflective states by adjusting heater power.
Deviations at lower heater powers arise from resistivity fluctuations due to heat dissipation and environmental influences.

Adjusting the power coupling ratios $k_1$ and $k_2$ in the directional couplers, along with applying appropriate heater powers (e.g., $\sim$600 mW as in Supplementary Figure \ref{Fig:TLM}(b)), enables opposite output behavior for the 1480 nm and 1550 nm inputs, effectively serving as a wavelength division multiplexer.

In the design of the broadband tunable loop mirrors for the EDWL device, two reflectors were engineered with distinct functionalities to optimize laser performance.
The output reflector provides tunable broadband reflectivity for lasing wavelengths, while the other reflects C- and L-band lasing and transmits 1480 nm Er pump light.
These functionalities were achieved by tailoring the coupling ratios $k_1$ and $k_2$ in the directional couplers.
Both mirrors are tunable via bias applied to the MZI arms.

Supplementary Figure \ref{Fig:TLM2} presents the experimental characterization of the tunable loop mirrors. 
By varying the heater voltage, transmission was measured and normalized to a reference waveguide, with polarization set to transverse electric (TE) mode.
The results demonstrate tunable broadband reflection (Supplementary Figure \ref{Fig:TLM2}(a)) for EDWL output power modulation, and tunable mode-selective reflection (Supplementary Figure \ref{Fig:TLM2}(b), heater power at 0 mW or $>$300 mW) for Er ion excitation and lasing mode reflection.

\section{Theoretical analysis of EDWL output power and linewidth}

\begin{figure*}[t!]
\centering
\includegraphics[width=0.9\textwidth]{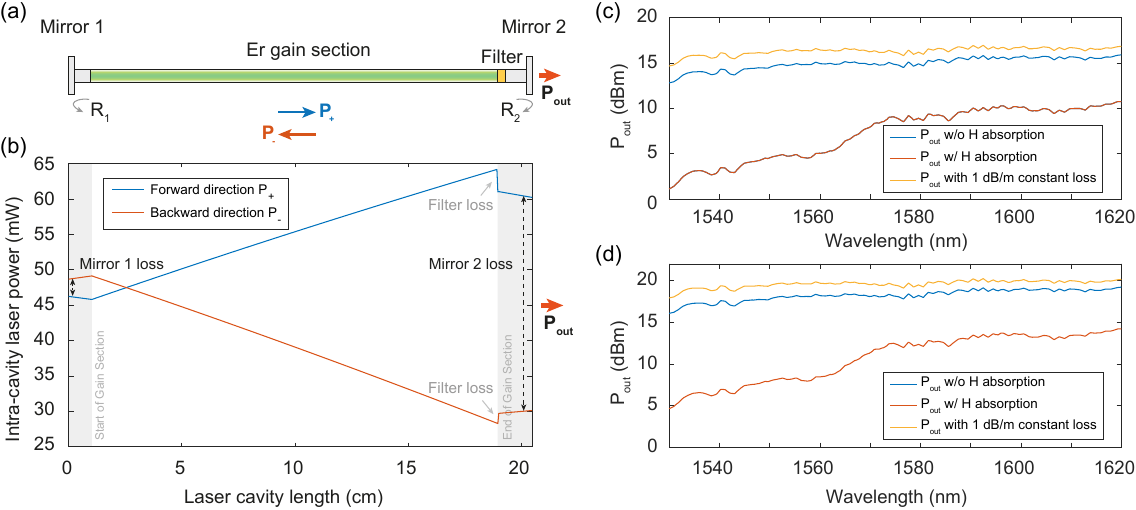}
\caption{
\footnotesize
\textbf{Simulation of intra-cavity power distribution and wavelength-dependent output power in an erbium-doped waveguide laser.}
(\textbf{a}) Schematic of the laser cavity, featuring two mirrors, an erbium-doped gain section, and a mode filter.
Simulated output power is collected from mirror 2.
(\textbf{b}) Simulated intra-cavity power distribution for forward ($P_{+}$) and backward ($P_{-}$) propagating light.
Losses due to mirror transmission and filtering are highlighted.
(\textbf{c}) Wavelength-dependent output power for an EDWL with an erbium ion concentration of 5.6$\times$10$^{26}$ m$^{-3}$ and an ion lifetime of 3.4 ms.
(\textbf{d}) Simulated output power for an EDWL with a higher ion concentration (9.35$\times$10$^{26}$ m$^{-3}$) and a shorter ion lifetime of 2.7 ms.
}
\label{Fig:Simulation_Pout}
\end{figure*}

This section investigates the theoretical performance of the erbium-doped waveguide laser (EDWL) by simulating the output power and fundamental linewidth using typical parameters presented in this work.

Supplementary Figure \ref{Fig:Simulation_Pout} illustrates the simulated intra-cavity power distribution together with the wavelength-dependent output power of the EDWL.
Fig. \ref{Fig:Simulation_Pout}(a) presents a schematic of the laser cavity, comprising a high-reflectivity back mirror (reflection $R_1$), an output coupling mirror (reflection $R_2$), an erbium-doped gain section, and a mode filter with transmission $T_\mathrm{f}$.
The rate equation governing the complex light field $A(t)$ within a single-mode laser cavity is expressed as:

\begin{equation}
\frac{n_\mathrm{g}L}{c}\cdot\frac{\partial A}{\partial t}=\left(\frac{g L}{2}-\frac{\alpha L}{2}\right) A(t)-\left[\ln \left(\frac{1}{\sqrt{R_1}}\right)+\ln \left(\frac{1}{\sqrt{R_2}}\right)+\ln \left(\frac{1}{\sqrt{T_\mathrm{f}}}\right)\right] A(t)
\end{equation}

where n$_\mathrm{g}$ is the group index, $L$ is the cavity round-trip length, $g$ is the gain coefficient (1/m), and $\alpha$ is the propagation loss (1/m).
The term n$_\mathrm{g} L / c$ represents the round-trip time, measured as 3.33 ns from the longitudinal mode spacing.
This exceeds the calculated value by 0.92 ns due to group delay from the Vernier filter microresonators.

Fig. \ref{Fig:Simulation_Pout}(b) illustrates the simulated intra-cavity power distribution for forward ($P_{+}$) and backward ($P_{-}$) propagating modes, where mirror transmissions and filter losses are highlighted.
Both modes experience loss throughout the cavity, with a round-trip loss coefficient of $e^{-\alpha L}$.
Amplification occurs in the Er-doped gain section with an effective gain coefficient $g_{\mathrm{eff}}=g_0/(1+P/P_{\mathrm{sat}})$.
Gain is modeled as $g_0=\sigma_\mathrm{e} N_2-\sigma_\mathrm{a} N_1$, where $N_2$ and $N_2$ are the population densities of the erbium ions at the excited state and the ground state, $\sigma_\mathrm{e}$ and $\sigma_\mathrm{a}$ are the Er emission and absorption cross-sections, using values measured in \cite{liu_photonic_2022}.
$P_{\mathrm{sat}}$ is the saturation power given by $P_{\mathrm {sat }}=\frac{h v A_{\mathrm {eff }}}{\tau (\sigma_\mathrm{a} + \sigma_\mathrm{e})} \cdot \frac{1}{\Gamma}$, where $\tau$ is the Er excited state lifetime \cite{singh2020towards}. In the simulation, mirror reflections are set to $R_1 = 0.95$ and $R_2 = 0.5$.
The gain and loss coefficients are 2.28 dB/cm and 1.98 dB/m, respectively, matching values measured in actual EDWL devices.
At the end of the Er-doped gain section, the forward mode encounters Vernier filter loss $(1-T_\mathrm{f})$ before being reflected by mirror 2 with reflection $R_2$.
The reflected light then propagates backward with initial power $R_2 \cdot P_{+}$.
Upon reaching mirror 1, it is reflected again to reinitiate $P_{+}$. The values of $P_{+}$ and $P_{-}$ are iteratively determined by applying boundary conditions until convergence.
The laser output from mirror 2 is given by $P_\mathrm{out} = (1-R_2) \cdot P_{+}$.
Reducing filter loss $T_\mathrm{f}$ and increasing mirror reflectivity $R_1$ can enhance the EDWL output power.

Fig. \ref{Fig:Simulation_Pout}(c)(d) present the theoretical wavelength-dependent output power in EDWLs with erbium ion concentrations of 5.6$\times$10$^{26}$ m$^{-3}$ and 9.35$\times$10$^{26}$ m$^{-3}$, and respective Er ion lifetimes of 3.4 ms and 2.7 ms.
Mirror reflections are set to $R_1=1$ and $R_2=0.5$, with filter losses neglected for maximum output power approximation.

\begin{figure*}[t!]
\centering
\includegraphics[width=0.9\textwidth]{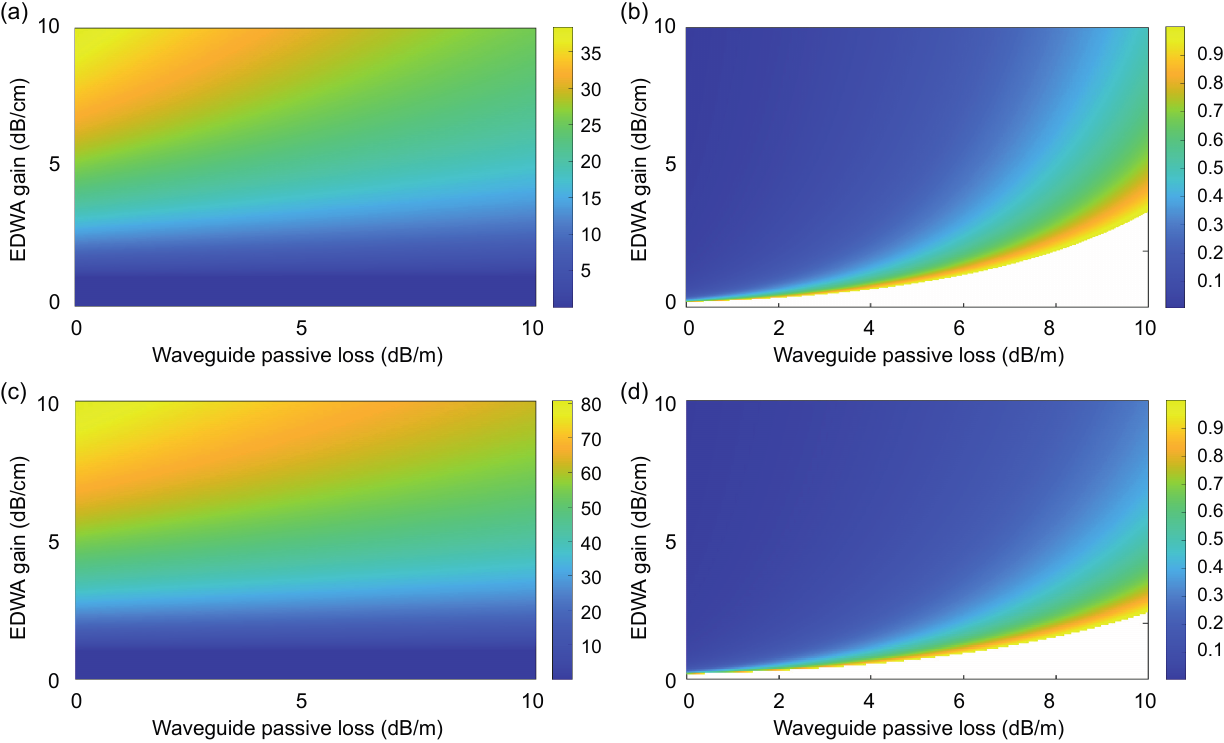}
\caption{
\footnotesize
\textbf{Simulated output power and Schawlow–Townes linewidth limit of EDWLs with varying erbium concentrations.}
Two EDWLs are compared: EDWL1 (Er concentration = 5.6$\times$10$^{26}$ m$^{-3}$, Er lifetime = 3.4 ms) and EDWL2 (9.35$\times$10$^{26}$ m$^{-3}$, 2.7 ms). 
(\textbf{a}) EDWL1 output power (mW) versus waveguide loss and amplifier gain.
(\textbf{b}) Schawlow–Townes linewidth limit (Hz) of EDWL1 as a function of waveguide loss and amplifier gain.
(\textbf{c}) EDWL2 output power (mW) versus waveguide loss and amplifier gain.
(\textbf{d}) Schawlow–Townes linewidth limit (Hz) of EDWL2 as a function of waveguide loss and amplifier gain.
}
\label{Fig:Simulation_gain_loss}
\end{figure*}

Fig. \ref{Fig:Simulation_Pout}(c) evaluates three loss conditions: Er-doped Si$_3$N$_4$ (Er:Si$_3$N$_4$) waveguides with low-loss SiO$_2$ cladding, SiO$_2$ cladding with absorptive O-H impurities, and wavelength-independent passive loss of 1 dB/m.
O-H bonds in the SiO$_2$ cladding of the fabricated EDWL, due to exposure to water in the ICPCVD chamber \cite{qiu_hydrogen_free_2023}, strongly absorb light at 200 THz, with an extended tail in the C-band \cite{afrin2013water}.
This absorption significantly reduces EDWL output power at shorter wavelengths, consistent with experimental findings in the main text.
The L-band EDWL output power matches the simulated value, indicating a smaller O-H absorption tail in real devices than initially modeled.

Simulations using Er:Si$_3$N$_4$ waveguides with absorption-free SiO$_2$ cladding show uniform output power across the C- and L-bands.
Despite stronger emission from Er ions in the C-band, higher saturation power and reduced propagation loss at longer wavelengths result in a relatively flat output power over the Er emission band.
Additionally, a wavelength-independent uniform loss of 1 dB/m further enhances EDWL power and output flatness.

In Fig. \ref{Fig:Simulation_Pout}(d), the same loss levels were applied to EDWL with a higher Er implantation dosage, resulting in an Er concentration of 9.35$\times$10$^{26}$ m$^{-3}$ and a lifetime of 2.7 ms, yielding increased EDWL output power.
However, higher erbium concentrations can reduce the excited state lifetime and potentially enhance re-absorption and cooperative upconversion effects, reducing efficiency and alternating the Er absorption and emission characteristics \cite{myslinski1997effects,dong2004concentration}.
This highlights the need for careful optimization of the doping concentration to balance the trade-offs between increased gain and potential loss mechanisms.

The influence of erbium ion concentration, waveguide loss, and amplifier gain on laser performance is examined in Supplementary Figure \ref{Fig:Simulation_gain_loss}, using the same algorithm and parameters as above.
Fig. \ref{Fig:Simulation_gain_loss}(a) shows output power (mW) dependence on waveguide passive loss (dB/m) and erbium-doped amplifier gain (dB/cm) at 1550 nm lasing.
Higher passive loss reduces output power, especially at low amplifier gain, underscoring the need to minimize passive losses for maximum efficiency.
Increasing amplifier gain can mitigate these losses and boost output power, but it also raises pump power requirements and introduces thermal management challenges.

Fig. \ref{Fig:Simulation_gain_loss}(b) illustrates the Schawlow–Townes laser linewidth limit (Hz) as a function of waveguide loss (dB/m) and amplifier gain (dB/cm).
The fundamental linewidth is given by $\Delta \nu=n_\mathrm{sp} \pi h \nu \kappa^2/P_{\mathrm{out}}$, 
$n_\mathrm{sp}=N_2/(N_2-N_1)$ is the spontaneous emission factor.
At thermal equilibrium, the populations in the excited state $N_2$ and the ground state $N_1$ follow the Boltzmann distribution: $N_2/N_1=e^{-\frac{h\Delta\nu}{kT}}$.
From \cite{mccumber1964einstein}, the cross-sections satisfy the relation: $\sigma_\mathrm{a}(\nu)=\sigma_\mathrm{e}(\nu) e^{\frac{h\Delta\nu}{k T}}$, at the same spatial eigenmode.
The population densities can be expressed as:

\begin{equation}
\begin{aligned}
& N_1=\frac{\sigma_\mathrm{a}}{\sigma_\mathrm{e}+\sigma_\mathrm{a}} N_0\\
& N_2=\frac{\sigma_\mathrm{e}}{\sigma_\mathrm{e}+\sigma_\mathrm{a}} N_0
\end{aligned}
\end{equation}
where $N_0$ is the total population of the implanted Er ions.
The spontaneous emission factor is calculated as $n_\mathrm{sp}=1.75$ for strong 1480 nm optical pumping.
The linewidth $\Delta \nu$ decreases with increasing amplifier gain, as higher gain reduces the relative noise contribution, while increased passive losses broaden the linewidth due to greater intra-cavity attenuation. Thus, minimizing passive losses and optimizing amplifier gain are key to achieving narrow linewidth for applications requiring high spectral purity.

Fig. \ref{Fig:Simulation_gain_loss}(c)(d) shows a similar study for the EDWL with a higher Er concentration, displaying comparable output power and linewidth trends with gain and loss, and achieving higher output power under the same waveguide conditions, consistent with Fig. \ref{Fig:Simulation_Pout}(c)(d).

\section{Frequency noise transduction from pump laser RIN}
\label{sec:RIN_trans}
\begin{figure*}[h!]
    \centering
    \includegraphics[width=0.9\textwidth]{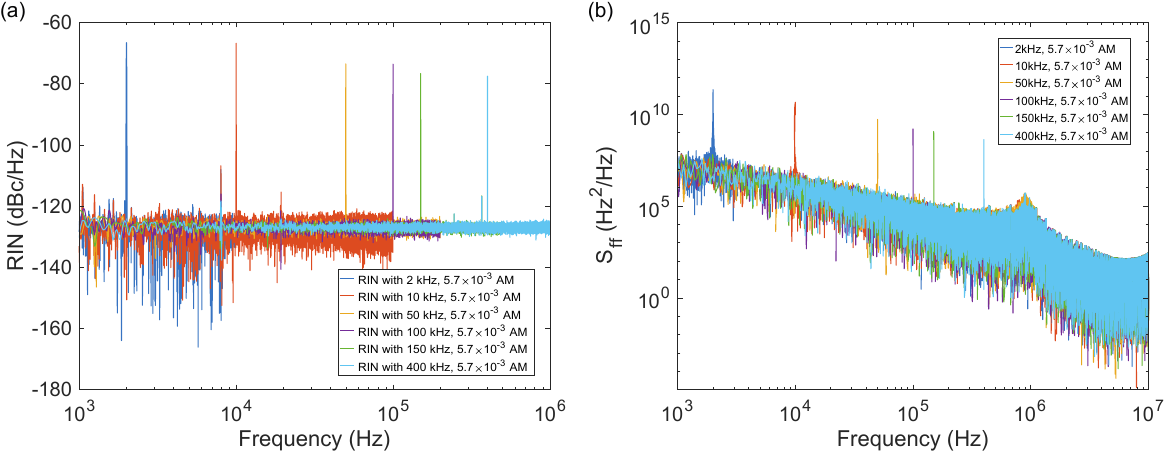}
    \caption{
    \footnotesize
    \textbf{Pump RIN and phase noise spectrum with modulation} (a) Measured pump RIN with amplitude modulation of $5.7\times10^{-3}$ applied to the drive current at frequencies of 2 kHz, 10 kHz, 50 kHz, 100 kHz, 150 kHz, and 400 kHz. (b) Corresponding measured laser FN obtained via delayed self-heterodyne interferometry (DSHI) at each modulation frequency.}
    \label{Fig:S2}
    \end{figure*}
    
We examine the contribution of pump laser RIN to the EDWL frequency noise.

We characterized the transduction from pump laser's intensity modulation \cite{lucas2020ultralow} to the phase modulation of the Vernier laser.
Using an arbitrary waveform generator, we applied a sinusoidal modulation to the drive current of the pump laser diode (LD), which had an average drive current of 1400 mA. With low modulation depth and within the modulation bandwidth, the relationship between current modulation and laser frequency noise was assumed to be  linear. This was verified by the absence of harmonic peaks on the RIN spectrum after applying the modulation.
The pump LD controller had a norminal modulation coefficient of 200~mA/V and a modulation bandwidth of 1.2 MHz. We set the peak-to-peak voltage of the sinusoidal signal at 40~mV, producing an amplitude modulation of approximately $5.7\times10^{-3}$ in the pump drive current.
        
\begin{figure*}[t!]
        \centering
        \includegraphics[width=0.9\textwidth]{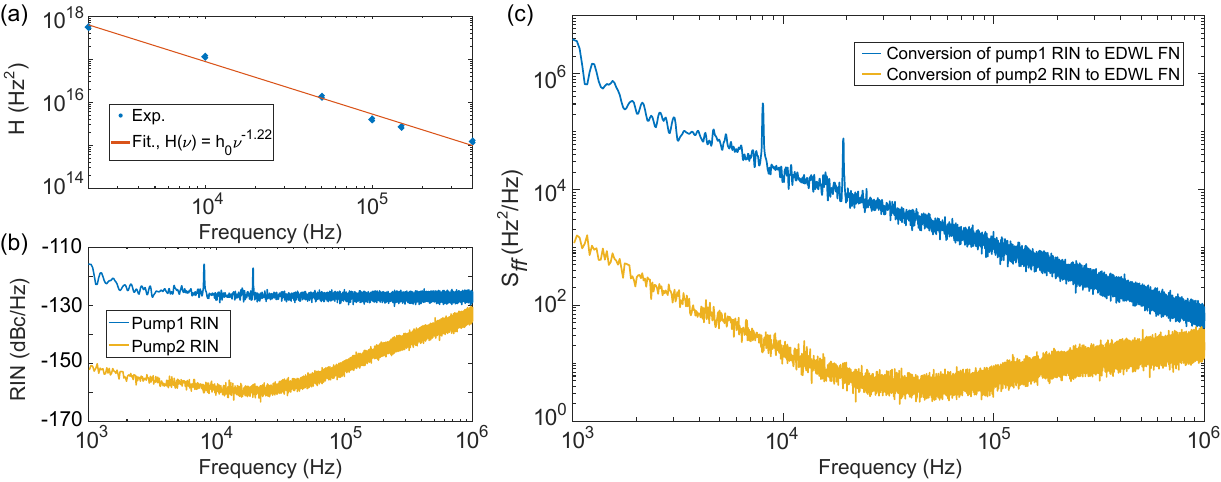}
        \caption{
        \footnotesize
        \textbf{Transduction function and the computed noise contribution} (a) Measured and fitted transduction function $H(\nu)$
        according to equation \ref{eq:transfer func}. (b) Measured relative intensity noise (RIN) of two different 1480~nm laser diodes: Pump 1 and Pump 2. (c) Calculated laser frequency noise (FN) transduced from the RIN of Pump 1 and Pump 2.
        }
        \label{Fig:S3}
        \end{figure*}

At a few modulation frequencies $\nu$ from 2 kHz to 400 kHz, we measured both the pump RIN and laser frequency noise (Figure \ref{Fig:S2}(a)(b)). The transduction function $H(\nu)$ is calculated as follows: 
\begin{equation}
    H(\nu)=\frac{\int_{\text {peak }} S_\mathrm{ff}(\nu') d \nu'}{\int_{\text {peak }} \text {RIN}(\nu') d \nu'}
    \label{eq:transfer func}
\end{equation}
where the integration is performed over the peaks at modulation frequency $\nu$ in the laser frequency noise power spectral density and pump RIN. The transduction function exhibits a linear relation on a log-log scale (Figure \ref{Fig:S3}(a)). 

As shown in Figure \ref{Fig:S3}(c), the transduced frequency noise was determined by multiplying the pump RIN without modulation (Figure \ref{Fig:S3}(b)) by the transduction function $H(\nu)$:
\begin{equation}
    S_\mathrm{ff}(\nu)=H(\nu)\cdot\text{RIN}(\nu)
    \label{eq:transducted noise}
\end{equation}

\section{Frequency noise transduction from thermal refractive noise in Vernier ring resonators}
\begin{figure*}[h!]
    \centering
    \includegraphics[width=0.4\textwidth]{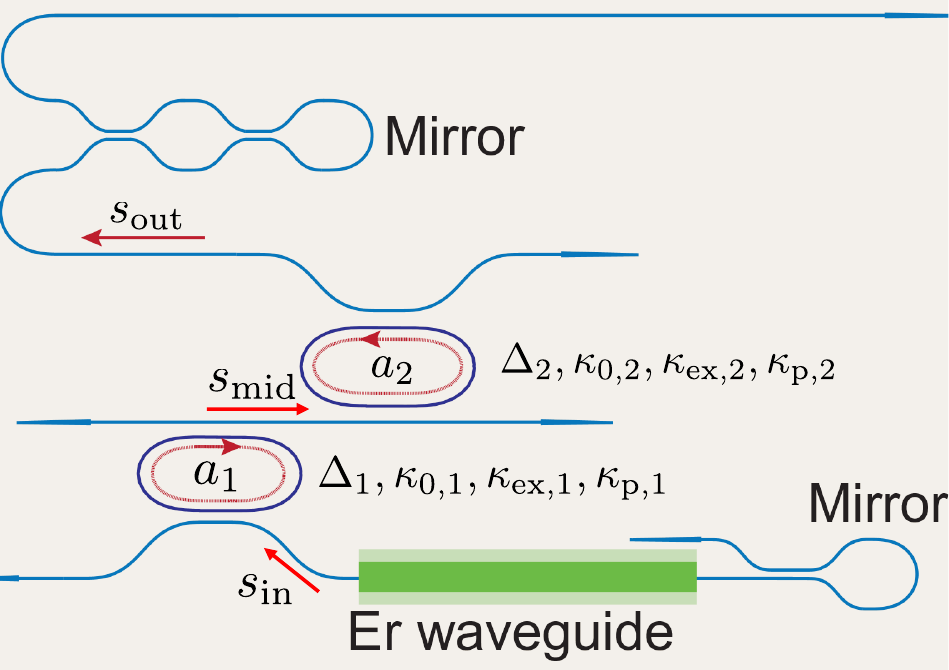}
    \caption{
    \footnotesize
    \textbf{Schematic illustrating the analysis of transduction from TRN in Vernier ring resonators to laser FN.}
    }
    \label{Fig:TRN transduction}
    \end{figure*}
To evaluate the contribution of thermal refractive noise (TRN) in Vernier ring resonators, we consider a double-resonator model, as illustrated in Figure \ref{Fig:TRN transduction}, and analyze its transmission characteristics.
The Langevin equations for the system can be written as follows: 
\begin{equation}
    \frac{da_1}{dt}=-i\Delta_1 a_1-\frac{\kappa_1}{2}a_1+\sqrt{\kappa_{\text{ex,1}}}s_{\text{in}}
    \label{eq:Langevin equation1}
\end{equation}
\begin{equation}
    \frac{da_2}{dt}=-i\Delta_2 a_2-\frac{\kappa_2}{2}a_2+\sqrt{\kappa_\mathrm{\text{ex,2}}}s_\mathrm{\text{mid}}
    \label{eq:Langevin equation2}
\end{equation}
and the coupling relation:
\begin{equation}
    s_{\text{mid}}=-\sqrt{\kappa_\mathrm{\text{ex,1}}}a_{1}, s_{\text{out}}=-\sqrt{\kappa_\mathrm{\text{ex,2}}}a_{2}
    \label{eq:Langevin equation3}
\end{equation}
where $s_{\text{in}}$, $s_{\text{mid}}$ and $s_{\text{out}}$ are the laser cavity modes at various points within the laser cavity, and $|s|^2$ corresponds to the photon flux at each respective position. For resonator j, $a_\mathrm{j}$ represents the resonator mode, with $|a_\mathrm{j}|^2$ denoting the photon number within the resonator. $\kappa_\mathrm{0,j}$, $\kappa_\mathrm{\text{ex},j}$, $\kappa_\mathrm{\text{p},j}$ and $\kappa_\mathrm{j}$ represent, respectively, the intrinsic loss, coupling strength to the fundamental mode, parasitic loss, and total loss. Finally, $\Delta_\mathrm{j}=\omega_\mathrm{j}-\omega$ denotes the detuning of the laser from the cavity mode, where $\omega$ is the lasing frequency.

Starting from equation \ref{eq:Langevin equation1}, \ref{eq:Langevin equation2} and \ref{eq:Langevin equation3}, we derive the relationship between $s_{\text{in}}$ and $s_{\text{out}}$:
\begin{equation}
    \Rightarrow s_{\text{out}}=\frac{\kappa_{\text{ex},1}}{i\Delta_1+\kappa_1/2}\frac{\kappa_{\text{ex},2}}{i\Delta_2+\kappa_2/2}s_{\text{in}}
    \label{eq:Langevin equation4}
\end{equation}
The phase relationship is given by:
\begin{equation}
    \varphi_{\text{vernier}}=-\tan^{-1}{\frac{2\Delta_1}{\kappa_1}}-\tan^{-1}{\frac{2\Delta_2}{\kappa_2}}
    \label{eq:Langevin equation5}
\end{equation}
For the longitudinal mode within the laser cavity, the roundtrip phase must be an integer multiple of $2\pi$:
\begin{equation}
    \varphi_{\text{rt}}=2\varphi_{\text{vernier}}+2\beta L_{\text{cav}}=2\pi m, m\in \mathbb{N}
    \label{eq:phase relation1}
\end{equation}
where $\beta=\frac{n_{\text{eff}}\omega}{c}$ is the effective propagation constant and $L_{\text{cav}}$ is the cavity length.
We introduce a perturbation to the roundtrip phase, and the resulting variance must be zero:
\begin{equation}
    \delta\varphi_{\text{rt}}=\frac{\partial\varphi_{\text{rt}}}{\partial\omega}\delta\omega+\frac{\partial\varphi_{\text{rt}}}{\partial\omega_1}\delta\omega_1+\frac{\partial\varphi_{\text{rt}}}{\partial\omega_2}\delta\omega_2=0
    \label{eq:phase relation2}
\end{equation}
The derivatives are expressed as follows:
\begin{equation}
    \frac{\partial\varphi_{\text{rt}}}{\partial\omega}=\frac{4/\kappa_1}{1+(2\Delta_1/\kappa_1)^2}+\frac{4/\kappa_2}{1+(2\Delta_2/\kappa_2)^2}+\frac{2L_{\text{cav}}}{c}n_\mathrm{g}
    \label{eq:phase relation3}
\end{equation}
\begin{equation}
    \frac{\partial\varphi_{\text{rt}}}{\partial\omega_1}=-\frac{4/\kappa_1}{1+(2\Delta_1/\kappa_1)^2},\ \frac{\partial\varphi_{\text{rt}}}{\partial\omega_2}=-\frac{4/\kappa_2}{1+(2\Delta_2/\kappa_2)^2}
    \label{eq:phase relation4}
\end{equation}
To maximize transduction and avoid underestimating the influence of TRN, we set $\Delta_1=\Delta_2=0$. For our Vernier ring design, we use $\kappa_1\approx\kappa_2\approx30~\kappa_0$, with $\kappa_0\approx~40~\text{MHz}\cdot2\pi$.
Given laser cavity length $L_{\text{cav}}\approx18~\text{cm}$ and effective group index $n_\mathrm{g}\approx1.8$, we obtain:
\begin{equation}
    \delta\omega=\frac{\delta\omega_1+\delta\omega_2}{2+L_{\text{cav}}\kappa_1 n_\mathrm{g}/2c}\approx\frac{\delta\omega_1+\delta\omega_2}{6}
    \label{eq:phase relation5}
\end{equation}

Therefore, the relationship between the laser FN and the frequency fluctuations of the Vernier rings due to TRN can be expressed as follows:
\begin{equation}
    S_\mathrm{f}=\frac{S_\mathrm{\delta f_1}+S_\mathrm{\delta f_2}}{36}\approx\frac{S_\mathrm{\delta f_1}}{18}
    \label{eq:phase relation6}
\end{equation}
where $S_\mathrm{\delta f_1}$ is the frequency noise power spectrum density of a single Vernier ring due to TRN. Here we approximate the two Vernier rings to have the same temperature, approximately 400 K, although this slightly overestimates the transduced FN, as in experiments one of the rings is maintained at room temperature.

\section{Analysis of pump laser dependent laser frequency noise}

\begin{figure*}[h!]
\centering
\includegraphics[width=0.7\textwidth]{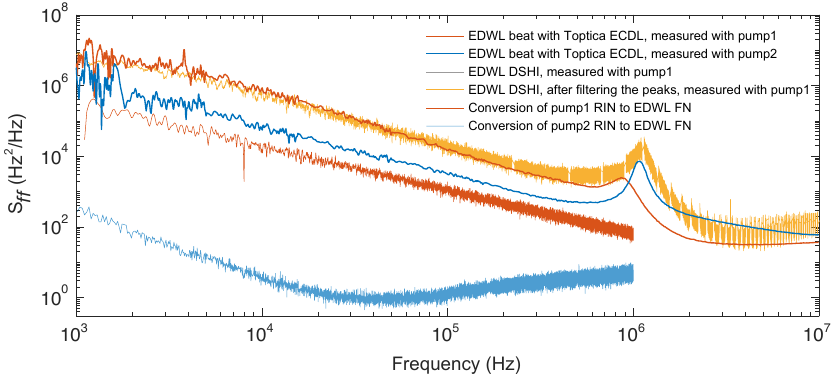}
\caption{
\footnotesize
\textbf{Impact of pump laser diode on EDWL frequency noise.}
The spectra include heterodyne beat measurements with an external cavity diode laser, when EDWL is pumped by two diodes with different RIN, delayed self-heterodyne measurements (yellow curve with MZI resonance peaks filtered), and transduced pump RIN to frequency noise for reference.
}
\label{Fig:pump_FN}
\end{figure*}

In Supplementary Section \ref{sec:RIN_trans}, we analyzed the impact of pump laser RIN on frequency noise in EDWL, observing that higher pump RIN leads to increased frequency noise.
In this section, we compare EDWL frequency noise obtained with different pump lasers and measurement techniques.

Supplementary Figure \ref{Fig:pump_FN} presents the EDWL frequency noise measured via heterodyne interferometry, when pumped with two diode lasers, along with delayed self-heterodyne measurements to determine the intrinsic frequency noise floor.
The light blue and red traces, shown up to 1 MHz, represent the transduced frequency noise from the pump laser RIN, as plotted in Supplementary figure \ref{Fig:S3}(c).
The blue and red EDWL noise traces display the frequency noise measured with each pump laser.
The low-RIN pump reduces EDWL frequency noise by nearly an order of magnitude at low offset frequencies, highlighting the impact of pump laser RIN on EDWL performance.
With the low-RIN pump, frequency noise is primarily limited by the thermal-refractive noise (TRN) of the microresonator in the Vernier filter.
The shift in relaxation oscillation peak positions is attributed to variations in intra-cavity laser power, influenced by the reflection conditions in the tunable loop mirror.

The yellow EDWL noise trace, from delayed self-heterodyne interferometry (DSHI), includes MZI transfer function resonances, which have been filtered to reveal the intrinsic laser frequency noise.
A similar noise floor to that observed in heterodyne beat spectroscopy with pump 1 is identified.

\section{Effect of heater resistivity drift on laser frequency stability}

\begin{figure*}[h!]
\centering
\includegraphics[width=0.8\textwidth]{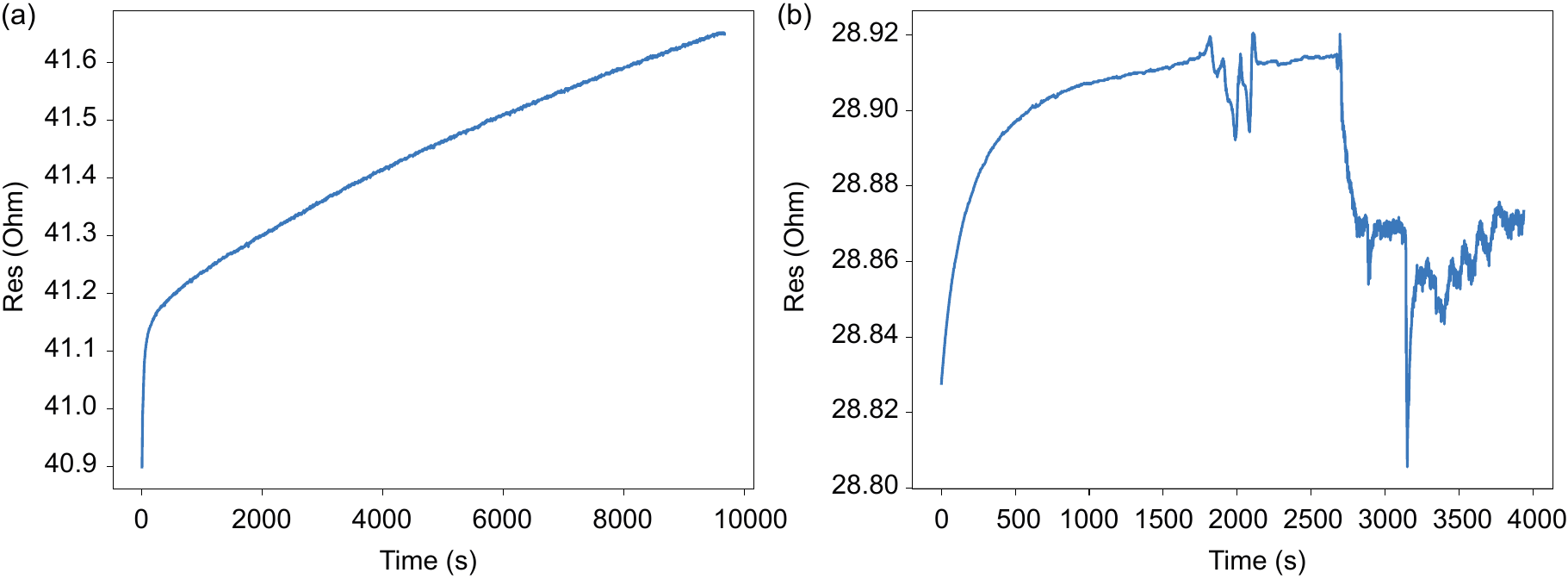}
\caption{
\footnotesize
\textbf{Micro-heater resistance variation over time.}
(\textbf{a}) Heater resistance drift in an isolated environment, illustrating a gradual change over time. 
(\textbf{b}) Heater resistance drift with isolation disruption at 2700 s, showing sudden changes from environmental factors.
}
\label{Fig:heater}
\end{figure*}

This section identifies micro-heater resistance drift as a source of EDWL frequency drift, influencing lasing frequency through resonance alignment in the Vernier filter and output power adjustment via tunable loop mirrors.

The micro-heaters, with a 5$\times$0.525 $\mu$m$^2$ cross-section, are fabricated through sequential DC sputtering of a 25 nm titanium adhesion layer and a 500 nm platinum layer onto the SiO$_2$ top cladding.
Heater are formed by direct laser writing onto a 3 $\mu$m AZ 10XT photoresist layer, followed by Argon ion beam etching.

After fabricating the micro-heaters, a constant voltage is applied, and the output current is monitored to calculate resistance drift as the voltage-to-current ratio.
Supplementary Figure \ref{Fig:heater}(a) shows heater resistance drift in an isolated environment (experimental setup enclosed with a cover) over 2.7 hours, exhibiting a continuous upward trend.
In contrast, Supplementary Figure \ref{Fig:heater}(b) highlights sudden changes and fluctuations after removing the cover at 2700 s, exposing the micro-heater to ambient conditions.

Resistance changes are common in semiconductor devices~\cite{colinge2005physics}, particularly in thin-film resistive materials like the micro-heaters used in EDWLs.
In metal conductors, gradual resistance changes can arise from oxidation~\cite{cabrera1949theory} and electromigration~\cite{tu2003recent} caused by constant current flow.
Environmental factors such as humidity, contamination, and airflow further exacerbate these effects.
Isolating the EDWL from the surrounding environment is therefore critical, and alternative actuators avoiding heat generation, for instance AlN or PZT may help mitigate laser frequency drift.

\clearpage

\bigskip
\bibliographystyle{apsrev4-2}
\bibliography{SI}